\documentclass[12pt]{article}
\usepackage[margin=1in]{geometry}

\title{Simultaneous Transformation and Rounding (STAR) Models for Integer-Valued Data}
\author{Daniel R. Kowal\thanks{Assistant Professor, Department of Statistics, Rice University, Houston, TX  (daniel.kowal@rice.edu).} \ and Antonio Canale\thanks{Assistant Professor, Department of Statistical Sciences, University of Padova, Padova, Italy.}}
\date{\today}

\usepackage{natbib}
\usepackage{graphicx}
\usepackage{chngcntr}

\usepackage{multirow,multicol}
\usepackage{amssymb,amsfonts,amsthm,amsmath} 
\usepackage{fullpage}
\usepackage{enumitem}
\usepackage{bm}
\usepackage{color}
\usepackage{hyperref}

\usepackage{setspace}
\newcommand{\STAR}{\textsc{star} }
\newcommand{\STARp}{\textsc{star}}
\newcommand{\R}{$\mathbb{R}$ }
\newcommand{\bart}{\textsc{bart} }
\newcommand{\bartp}{\textsc{bart}}

\theoremstyle{plain}   \newtheorem{lemma}{Lemma}

\begin{document}

\maketitle

\vspace{-12mm}
\begin{abstract}
We propose a simple yet powerful framework for modeling integer-valued data, such as counts, scores, and rounded data. The data-generating process is defined by Simultaneously Transforming and Rounding (\STARp) a continuous-valued process, which produces a flexible family of integer-valued distributions capable of modeling zero-inflation, bounded or censored data, and over- or underdispersion. The transformation is modeled as unknown for greater distributional flexibility, while the rounding operation ensures a coherent integer-valued data-generating process. An efficient MCMC algorithm is developed for posterior inference and provides a mechanism for adaptation of successful Bayesian models and algorithms for continuous data to the integer-valued data setting. Using the \STAR framework, we design a new Bayesian Additive Regression Tree (\bartp) model for integer-valued data, which demonstrates impressive predictive distribution accuracy for both synthetic data and a large healthcare utilization dataset. For interpretable regression-based inference, we develop a \STAR additive model, which offers greater flexibility and scalability than existing integer-valued models.  The \STAR additive model is applied to study the recent decline in Amazon river dolphins.
\end{abstract}




\section{Introduction}
A challenging scenario for prediction and inference occurs when the outcome variables are integer-valued, such as counts, (test) scores, or rounded data. Integer-valued data are ubiquitous in many fields, including epidemiology \citep{osthus2018dynamic,kowal2019measles},  ecology \citep{dorazio2005improving},  and insurance \citep{bening2012generalized},  among many others \citep{cameron2013regression}.  Counts often serve as an indicator of demand, such as the demand for medical services \citep{deb1997demand}, emergency medical services \citep{matteson2011forecasting}, and call center access \citep{shen2008interday}.
Integer-valued data are discrete data, and exhibit a variety of complex distributional features including zero-inflation, skewness, over- or underdispersion, and in some cases may be bounded or censored. Consequently, prediction and modeling of integer-valued data---in the presence of predictors, over time intervals, and across spatial locations---remains a significant challenge. 

The most widely-used models for integer-valued data build upon the Poisson distribution. 
However, the  limitations of the Poisson distribution are well-known: the distribution is not sufficiently flexible in practice and cannot account for zero-inflation or over- and underdispersion. A common strategy is to generalize the Poisson model by introducing additional parameters, such as the quasi-Poisson \citep{mccullagh1989generalized}, negative-binomial \citep{hilbe2011negative}, zero-inflated Poisson or negative-binomial \citep{cameron2013regression,neelon2019bayesian}, lognormal Poisson \citep{zhou2012lognormal}, restricted generalized Poisson \citep{famoye1993restricted}, and Conway-Maxwell Poisson models \citep{shmueli2005useful,lord2008application,sellers2010flexible}. A fundamental limitation of these approaches is that the additional parameters can introduce formidable challenges for estimation and computational scalability, especially in conjunction with regression, temporal, or spatial models.



In practice, however, it is exceedingly common for the discrete nature of the data to be ignored. 
Practitioners often log- or square-root-transform the observed integer-valued data and subsequently apply methods designed for continuous or Gaussian data. 
However, transformations to Gaussianity are ineffective for small counts \citep{warton2018you}, while log-transformations introduce difficulties in the presence of zeros \citep{o2010not}. More broadly, these approaches are not well-defined for integer-valued data: the data-generating process for a (transformed) Gaussian model cannot produce discrete data, which immediately amplifies model misspecification, limits interpretability, and undermines the reliability of inference and predictive distributions.  

To address these challenges, we propose a coherent modeling framework for integer-valued data. The process is defined by \emph{simultaneously transforming and rounding} (\STARp) a continuous-valued process. 
First, a \emph{continuous-valued process} is specified to model the dependence between (latent) variables. We focus on conditionally Gaussian regression models, but the \STAR framework applies more broadly. Second, the latent variables are \emph{transformed} for greater distributional flexibility. While the transformation may be specified in advance, such as logarithmic or square-root, we develop both parametric and nonparametric approaches to learn the transformation from the data, which improves predictive accuracy. 
 Lastly, the transformed latent variables are filtered through a \emph{rounding operator} mapping them to the (nonnegative) integers. This construction is inspired by the popular approach of transforming count data and applying Gaussian models, 
 yet produces a mathematically consistent and well-defined integer-valued process. Importantly, we show that \STAR processes are not merely \emph{valid} integer-valued distributions, but also \emph{flexible} integer-valued  distributions, and can account for zero-inflation, bounded or censored data, and over- or underdispersion. 



Another major benefit of \STAR is its computational modularity: using a simple and efficient data augmentation technique, existing computational tools for Bayesian inference under continuous data models can be used for Bayesian inference under \STAR models. As a result, \STAR provides a cohesive framework for seamlessly adapting state-of-the-art continuous data models and algorithms to the integer-valued data setting.  Using the \STAR framework, we design---among others---a new Bayesian Additive Regression Tree (\bartp) model for integer-valued data. The resulting \bartp-\STAR model combines the integer-valued distributional flexibility provided by \STAR with the predictive and computational advantages inherent to \bartp. For synthetic data and a large healthcare utilization dataset (Section~\ref{sims}-\ref{sec-dolphins}), the predictive performance of \bartp-\STAR far exceeds that of competing methods which do not include \emph{both} transformation and rounding in terms of out-of-sample predictive accuracy, model adequacy, and computational scalability. 

We also apply \STAR to study the recent decline in the tucuxi dolphin population, which inhabit the Amazon River.  Using field survey data conducted by \cite{da2018both} from 1994 to 2017, we develop a \STAR additive model for the number of observed tucuxi dolphins, which includes a smooth regression term for important predictor variables such as the year, day-of-year, and water level. The \STAR additive model is interpretable yet flexible, and demonstrates favorable performance in model fit and computational efficiency relative to existing integer-valued models.

The remainder of the paper is organized as follows. Section~\ref{star} introduces the \STAR framework, develops models for the unknown transformation, describes important properties, and discusses computational  details for posterior inference. Section~\ref{model-ex} provides example \STAR models, 
which are applied to simulated data in Section~\ref{sims} and real data in Sections~\ref{sec-nmes}~and~\ref{sec-dolphins}.
Section~\ref{discuss} concludes. Additional simulation results and empirical comparisons are in the Appendix. Methods are implemented in the \texttt{R} package \texttt{r\STARp} available on GitHub.

\section{Simultaneously transforming and rounding}\label{star}
Consider a count-valued stochastic process $y\!:\mathcal{X}\to \mathcal{N}$, where $\mathcal{X}$ may correspond to predictors, times, or spatial locations and $\mathcal{N} = \{0,\ldots,\infty\}$. Although we focus on the nonnegative integers, our procedure may be trivially modified for integer-valued data and rounded data. Our goal is to construct a joint probability distribution for $y$ that simultaneously builds upon successful approaches for continuous stochastic processes (observed on $\mathbb{R}$ or $\mathbb{R}^+$), yet produced a flexible and well-defined distribution on $\mathcal{N}$. 

To this end, we first introduce continuous-valued process $y^*\!:\mathcal{X}\to\mathcal{T}$, $\mathcal{T} \subseteq \mathbb{R}$ related to  the observed count-valued data $y$ via 
\begin{equation}\label{round}
y = h(y^*), 
\end{equation}
where $h\!: \mathcal{T}\to \mathcal{N}$ is a \emph{rounding} operator that sets $y(x) = j$ when $y^*(x) \in \mathcal{A}_j$ and $\{\mathcal{A}_j\}_{j=0}^\infty$ is a known partition of $\mathcal{T}$. 
For example, we may use the floor function defined by $\mathcal{A}_j = [a_j, a_{j+1}) = [j, j+1)$ for $j \in \mathcal{N}$; modifications are available for zero-inflated, bounded, or censored data (Section~\ref{sec-properties}).
The process $y^*$ operates as a continuous proxy for the observed counts $y$,  which is more convenient for modeling, yet has a simple mapping to the observable  data in \eqref{round}. Naturally, the properties of the count-valued process $y$ will be determined by the rounding operator $h$ and the distribution of the continuous-valued process $y^*$. 

We propose to induce a distribution on $y^*$ by \emph{transforming} $y^*$ and specifying a distribution $\Pi_\theta$ on the transformed scale:
\begin{equation}\label{transform}
g(y^*) = z^*, \quad z^*  \sim  \Pi_\theta,
\end{equation}
where $g\!:\mathcal{T} \to \mathbb{R}$ is a strictly monotone function. Model \eqref{transform} is inspired by the common practice of transforming count data prior to application of continuous (Gaussian) models. However, the \STAR framework of \eqref{round}-\eqref{transform} defines an integer-valued process for $y$, in which the transformation $g$ may be modeled as unknown for greater distributional flexibility.




While \STAR is sufficiently general to incorporate any continuous family of stochastic process $\Pi_\theta$ for the latent $z^*$, an important special case of \eqref{transform} is the conditionally Gaussian regression model:
\begin{equation}\label{simpleGP}
z^*(x) = \mu(x) + \epsilon(x), \quad \epsilon(x) \stackrel{indep}{\sim} N(0, \sigma^2(x)),
\end{equation}
where $\mu(x)$ is the conditional expectation of $z^*(x)$ and the errors $\epsilon(x)$ are independent but possibly heteroscedastic, conditional on $x \in \mathcal{X}$. 
Examples of \eqref{simpleGP} include linear and additive models (Section~\ref{sec-am}) and \bart (Section~\ref{sec-bart}), with extensions for mixed effects models, spatio-temporal models, dynamic linear models, and factor models, among others.




Rounding of a continuous process has appeared previously in the literature  \citep{canale2011bayesian,canale2013nonparametric}.  Our key innovation is the coupling of the \emph{transformation} \eqref{transform} with the regression model \eqref{simpleGP}. The transformation $g$, which we model as unknown, endows the integer-valued process $y$ with greater distributional flexibility, yet leaves model \eqref{simpleGP} unchanged. This construction allows seamless integration of Bayesian models \emph{and} algorithms for continuous data of the common form \eqref{simpleGP}  into the integer-valued \STAR framework, with efficient posterior inference available via a general MCMC algorithm (Section~\ref{mcmc}). As demonstrated extensively in the simulations and applications (Sections~\ref{sims}-\ref{sec-dolphins}), models that fail to include \emph{both} rounding and transformation cannot match the predictive performance of \STAR models.




The distribution of $y$ is completely determined by the rounding operator $h$, the transformation $g$, and the distribution $\Pi_\theta$. Specifically, the probability mass associated to $y(x) = j$ for each integer $j\in \mathcal N$ is 
\begin{equation}\label{pmf}
\mathbb{P}\{y(x) = j\} = \mathbb{P}\left\{y^*(x) \in \mathcal{A}_j\right\} = \mathbb{P}\left\{z^*(x) \in g(\mathcal{A}_j)\right\}. 
\end{equation}
The distribution of $z^*$ is given by $\Pi_\theta$, while $g(\mathcal{A}_j)$ is determined by the transformation $g$ and the rounding operator $h$. For model \eqref{simpleGP} and  $\mathcal{A}_j = [a_j, a_{j+1})$, 
\eqref{pmf} simplifies to  
\begin{equation}\label{simpleDist}
\mathbb{P}\{y(x) = j\} = \Phi\left( \frac{g(a_{j+1}) - \mu(x)}{\sigma(x)}\right) -  \Phi\left( \frac{g(a_{j}) - \mu(x)}{\sigma(x)}\right).
\end{equation}


The distribution in \eqref{simpleDist} is related to, yet distinct from, ordinal regression \citep{mccullagh1980regression}. In ordinal regression, each term $g(a_j)$ in \eqref{simpleDist} is replaced by an unknown latent threshold, say $\omega_j$, with an ordering constraint $\omega_j \le \omega_{j+1}$ for all $j$. However, the latent thresholds $\omega_j$ are based only on the \emph{ranks} of the observed data, and therefore ignore the information contained in the numeric values of the observed counts. Furthermore, 
since each threshold $\omega_j$ is unknown, ordinal regression introduces a new parameter for each unique data value, and therefore produces a heavily-parametrized model that is challenging to estimate. By comparison, \STAR is substantially more parsimonious: if $g$ is known, no new parameters are needed, while if $g$ is unknown, only a small number of parameters are needed (see Section~\ref{sec-transform}). 


 \STAR is fundamentally different from simply rounding the predictions from a continuous data model $\Pi_\theta$. \emph{Post hoc} rounding ignores the discrete nature of the data in model-fitting, and consequently introduces a disconnect between the \emph{fitted model} and the model used for \emph{prediction}. \STAR clearly avoids this issue, and maintains the benefits of using well-known models for continuous data while producing a coherent integer-valued predictive distribution.

\subsection{The transformation $g$}\label{sec-transform}
The transformation $g$ is a crucial component of \STARp. When $g(t) = t$ and $z^*$  is a draw from a Gaussian process,  \STAR simplifies to \cite{canale2013nonparametric}. However, the identity transformation is suboptimal in many cases (see Sections~\ref{sec-nmes}~and~\ref{sec-dolphins}). The popularity of log-linear models for count data, especially Poisson and negative-binomial models, suggests that regression effects $\mu(x)$ are often multiplicative for count data, and that the log-transformation $g(t) = \log(t)$ may be preferable for many applications. Similarly, the square-root transformation $g(t) = \sqrt{t}$ is the variance-stabilizing transformation of the Poisson distribution, and therefore is a common choice in applications of Gaussian methods to transformed count data. Empirically, the simulation studies and real data analyses in Sections~\ref{sims}-\ref{sec-dolphins} demonstrate that the transformation $g$ provides substantial improvements in modeling flexibility and accuracy relative to an untransformed approach.

When $g$ is fixed and known, 
the only unknowns are the parameters $\theta$ in the distribution $\Pi_\theta$ of the latent data $z^*$ in \eqref{transform}. Relative to the analogous transformed continuous-valued model, say $g(y) \sim \Pi_\theta$, the number of parameters is the same, yet \STAR produces a coherent integer-valued process.
 A fixed transformation $g$ shares some characteristics with the link function of a generalized linear model (GLM)  \citep{mccullagh1989generalized}. For GLMs, the link function maps the expectation of an exponential family distribution to $\mathbb{R}$, which is modeled using a linear predictor. By comparison, \STAR maps the continuous-valued $y^*$ to \R and under \eqref{simpleGP} models $\mathbb{E}[g\{y^*(x)\} | x] = \mathbb{E}\{z^*(x) | x\} = \mu(x)$. 

For general application of \STARp, pre-specification of a transformation $g$ is restrictive. By allowing the data to inform $g$, the implied distribution for $y$ becomes more flexible, and the risk of model misspecification is lessened. For GLMs,  \cite{mallick1994generalized} similarly relax the assumption of a known link function, adopting a nonparametric approach. For \STARp, we require that the functions $g$ satisfy the following properties: (i) \emph{monotonicity}, which preserves the ordering of the observed integers in the transformed latent space; (ii) \emph{smoothness}, which provides regularization by encouraging information-sharing among nearby values; and (iii) \emph{shrinkage} toward a pre-specified transformation, such as log or square-root. 

A natural parametric specification for $g$ satisfying the aforementioned criteria  is the (signed) \emph{Box-Cox} transformation \citep{box1964analysis}: 
\begin{equation}\label{box-cox}
g(t ; \lambda) = 
\{\mbox{sgn}(t)|t|^\lambda - 1\}/\lambda, \quad \lambda > 0 
\end{equation}
with $g(t; \lambda =0) = \log(t)$. Box-Cox functions are a popular choice for transforming continuous data towards Gaussianity, which in the present setting is similar to \eqref{transform} when $\Pi_\theta$ is Gaussian. Important special cases of \eqref{box-cox} include the (shifted) identity transformation $g(t; \lambda = 1) = t-1$, the (shifted and scaled) square-root transformation $g(t; \lambda = 1/2) = 2\sqrt{|t|} - 2$, and the log-transformation. 
To learn the shape of the transformation, we place a prior on $\lambda$: we recommend $\lambda \sim N(1/2, 1)$ truncated to $[0,3]$, which shrinks $g$ toward the (shifted and scaled) square-root transformation.

For additional flexibility, we also consider fully nonparametric specification for $g$. Consider an I-spline basis expansion  \citep{ramsay1988monotone} for $g$:
\begin{equation}\label{nonparag}
g(t) =  b_I'(t)  \gamma,
\end{equation}
where $ b_I$ is an $L$-dimensional vector of I-spline basis functions and $ \gamma$ are the unknown basis coefficients.  Since each I-spline basis function is monotone increasing, we ensure monotonicity of $g$ by restricting the elements of $\gamma$ to be positive. 

We propose a prior for $\gamma$ in \eqref{nonparag} that simultaneously enforces monotonicity, smoothness, and shrinkage toward a pre-specified transformation. However, care must be taken to ensure identifiability of the \STAR model and retain interpretability of the parameters $\theta$ in $\Pi_\theta$. For model \eqref{simpleGP}, arbitrary shifting and scaling of $g$ can be matched by shifting and scaling of $\mu$ and $\sigma$. The parametric transformation \eqref{box-cox} preserves identifiability: $g(1, \lambda) = 0$  for all $\lambda$ (shift constraint) and the prior on $\lambda$ is weakly informative (scale constraint). For nonparametric $g$ in \eqref{nonparag}, we resolve the identifiability issue by fixing $g(0) = 0$ (shift constraint), which is satisfied automatically due to the I-spline construction, and $\lim_{t\rightarrow\infty} g(t) = 1$ (scale constraint), which is enforced by constraining $\sum_{\ell=1}^L \gamma_\ell = 1$. Specifically, let
\begin{equation}\label{prior-gamma}
 \gamma_\ell = \tilde \gamma_\ell/\sum_{k = 1}^L \tilde \gamma_k, \quad \tilde\gamma_\ell \stackrel{indep}{\sim}N_+(\mu_{\gamma_\ell}, \sigma_\gamma^2), \quad \ell = 1,\ldots,L,
 \end{equation} 
where $N_+$ is the half-normal distribution. Clearly, $\gamma_\ell > 0$ for each $\ell$ and $\sum_{\ell=1}^L \gamma_\ell = 1$, which guarantees monotonicity of $g$ and preserves identifiability of model \eqref{simpleGP}. We select the prior mean $ \mu_\gamma = (\mu_{\gamma_1},\ldots, \mu_{\gamma_L})'$ such that $g(t)$ is \emph{a priori} centered around a parametric function of interest, such as \eqref{box-cox} with fixed $\lambda = \lambda_0$, and model $\sigma_{\gamma}^2$  with an inverse-Gamma prior to allow the data to determine the amount of shrinkage toward the parametric function of interest.  Let $\bm t_g = (0,1,\ldots, a_{\max y_i + 1}\})'$ and let $ B_I$ be the I-spline basis evaluated at $\bm t_g$, so $g(\bm t_g) =  B_I  \gamma$. We solve $ \tilde\mu_\gamma = \arg\min_{ \mu_\gamma} || g(\bm t_g; \lambda_0) -   B_I  \mu_\gamma||^2$ subject to $\tilde\mu_{\gamma_\ell} > 0$ for $\ell=1,\ldots,L$, which is a one-time cost, 
and normalize $\mu_{\gamma_\ell} = \tilde\mu_{\gamma_\ell}/\sum_{k=1}^L \tilde\mu_{\gamma_k}$. In the simulations and applications of Sections~\ref{sims}-\ref{sec-dolphins}, we fix $\lambda_0 = 1/2$ and model $\sigma_{\gamma}^{-2} \sim\mbox{Gamma}(0.001, 0.001)$.

We use quadratic I-splines with $L = 2 + \min\{(\mbox{\# unique } y_i)/4, 10\}$ knots, implemented using the  \texttt{splines2} package in \texttt{R} \citep{splines2}.  Boundary knots are placed at zero and $\max \{y_i\}$, while the $L-2$ interior knots selected using the sample quantiles of $\{y_i\}$ excluding zero, one, and  $\max \{y_i\}$, with an interior knot placed at one to improve distributional flexibility near zero. Since \cite{ramsay1988monotone} use $L=3$ or $L=5$ in all monotone spline examples, a small number of knots may be adequate in many cases.

\subsection{Model properties}\label{sec-properties}
By design, \STAR builds upon models for continuous data, such as those in Section~\ref{model-ex}, and adapts them for integer-valued data. Yet \STAR is not merely a mechanism for producing valid integer-valued processes: \STAR also provides important distributional properties for modeling integer-valued data in practice. By careful selection of the rounding operator $h$ and the transformation $g$, \STAR provides the capability to model zero-inflation, bounded or censored data, and over- or underdispersion. 

In applications with count data, it is common to observe an abundance of zeros, $y(x) = 0$.  \STAR can be parametrized such that zero counts occur whenever 
$z^*(x)$ is negative:
\begin{lemma}[Zero-inflation]\label{lem:zeroinflation}
For any \STAR model with $g(\mathcal{A}_0) = (-\infty, 0)$, we have  (i) $y(x) = 0$ if and only if $z^*(x) < 0$ and (ii) $\mathbb{P}\{y(x) = 0\} = \mathbb{P}\{z^*(x) \leq 0\}.$
\end{lemma}
Lemma \ref{lem:zeroinflation} is valid for known or unknown transformations, and is easily satisfied for \eqref{box-cox} letting $\mathcal{A}_0 = (a_0, a_1) = (-\infty, 1)$ for $\lambda \ne 0$ and $\mathcal{A}_0 = (a_0, a_1) = (0, 1)$ for $\lambda = 0$. For model \eqref{simpleGP}, $\Pi_\theta$ is conditionally Gaussian, which may place substantial prior mass on $z^*(x) < 0$ and thus  $y(x) = 0$. Therefore, \STAR has a built-in and interpretable mechanism for handling zero counts, and does not require the addition of an artificial constant to the transformation, such as $\log(y+1)$. Furthermore, dependence among zero values is implicit in the model: $\mathbb{P}\{y(x) = 0, y(x') = 0\} = \mathbb{P}\{z^*(x) < 0, z^*(x') < 0\}$ depends on the joint distribution of $(z^*(x), z^*(x'))$, which is modeled by $\Pi_\theta$.


Another common characteristic of count-valued data is a deterministic upper bound $K$. For instance, if $y$ counts the number of days on which an event occurred in a given year,  then $y(x) \in \{0,1,\ldots,K\}$ and $K = 365$. \STAR can easily incorporate this information into the distribution for $y$ as formalized in the next lemma.
\begin{lemma}[Bounded observations]\label{bounded}
For any \STAR model, letting $g(\mathcal{A}_K) = [g(a_K), \infty)$ implies  $\mathbb{P}\{y(x) \le K\} = 1$. 
\end{lemma}
The boundedness constraints in Lemma~\ref{bounded} are compatible with any choice of (unconstrained) continuous-valued model \eqref{transform} and do not require modification of the algorithms for estimation and inference in Section~\ref{mcmc}. 
Similar to the case of zero values in Lemma~\ref{lem:zeroinflation}, \STAR allows for dependence among $y$ values that attain the upper bound:  $\mathbb{P}\{y(x) = K, y(x') = K\} = \mathbb{P}\{z^*(x) \ge g(a_K), z^*(x') \ge g(a_K)\}$, which again is modeled by \eqref{transform} or \eqref{simpleGP}. When $K=1$, the \STAR model \eqref{simpleGP} with $\mu(x) = x'\beta$ and $\sigma^2(x) = 1$ simplifies to probit regression. 

 
Interestingly, the construction in Lemma~\ref{bounded} is coherent under right-censoring, which occurs when an observed count value of $K$ implies that $y(x) \ge K$. Right-censoring is common in surveys, where large values are often grouped together. The following lemma formalizes the properties of \STAR subject to right-censoring of the observations.
\begin{lemma}[Right-censoring]\label{censor}
Let $\STAR(h, g, \Pi_\theta)$ denote model \eqref{round}-\eqref{transform} with rounding operator $h$, transformation $g$, and latent distribution $\Pi_\theta$. 
For right-censored observations $y_c(x) = \min\{y(x), K\}$ with $y \sim \STAR(h, g, \Pi_\theta)$ and $y_c \sim \STAR(h', g, \Pi_\theta)$ such that $h$ and $h'$ satisfy $a_K = a'_K$ and  $\mathcal{A}'_K = [a_K, \infty)$, we have 
$\mathbb{P}\{y(x) \ge K\} = \mathbb{P}\{z^*(x) \ge g(a_K)\} = \mathbb{P}\{y_c(x) = K\}$. 
\end{lemma}
For right-censored data, the likelihood includes terms of the form $\mathbb{P}\{y(x) \ge K\}$ for censored observations. Lemma~\ref{censor} shows that the censored likelihood terms under a \STAR model for $y$ are equivalent to the non-censored likelihood terms $\mathbb{P}\{y_c(x) = K\}$ under a \STAR model for $y_c = \min\{y(x), K\}$ with $\mathcal{A}'_K = [a_K, \infty)$. Remarkably, \STAR preserves the correct right-censored likelihood for $y$ by directly modeling the observed counts $y_c$ and setting $\mathcal{A}_K' = [a_K, \infty)$, with no further modifications needed for the model specification or estimation procedure. By comparison, common parametric approaches for modeling count data, such as the Poisson model and its generalizations, require careful modifications of the likelihood and tailored algorithms for estimation and inference in the case of right-censoring. 
Naturally, a similar approach is available for left-censoring.



\begin{figure}[h!]
\begin{center}
\includegraphics[width=1\textwidth]{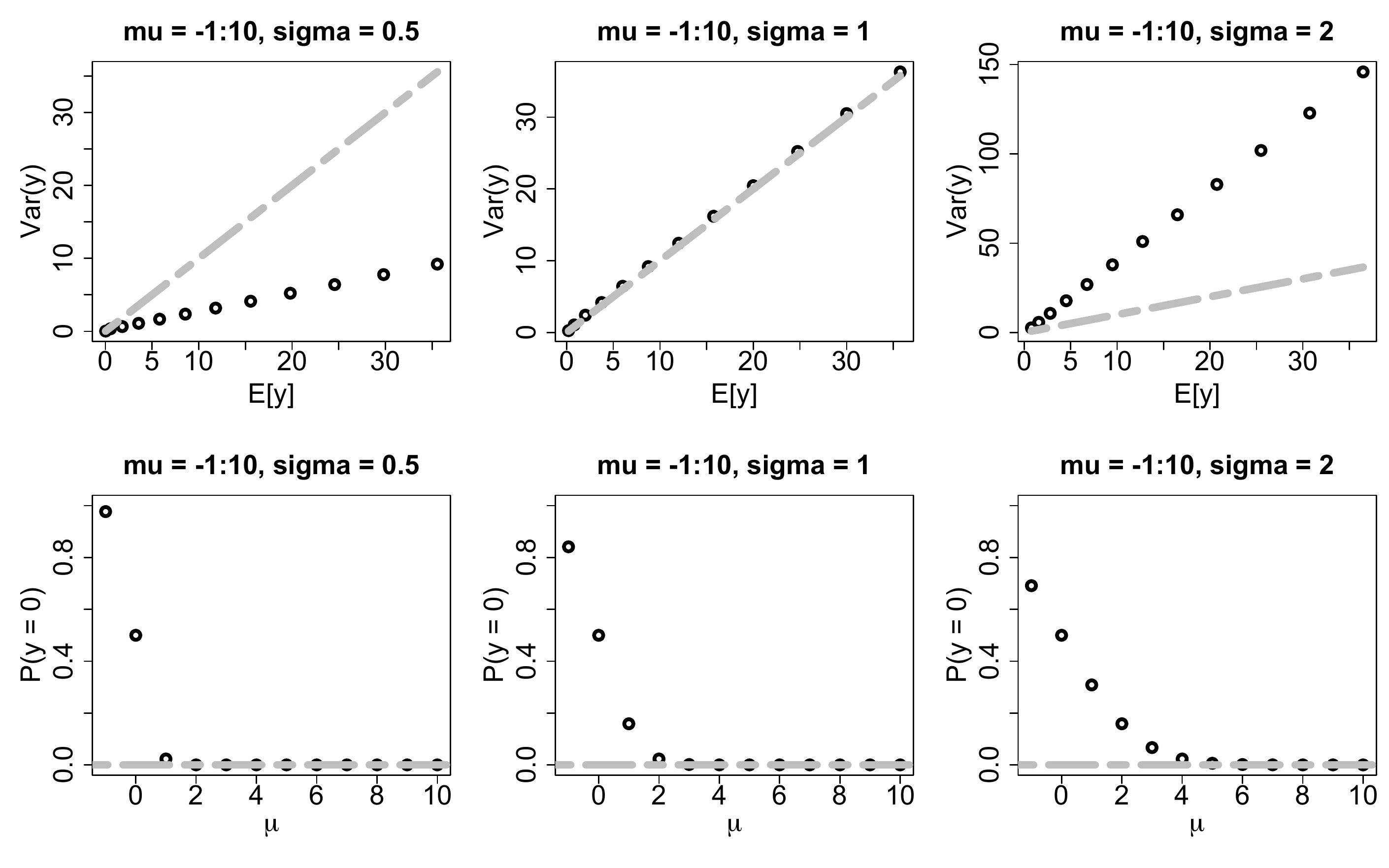}
\caption{\small $\mathbb{E}[y]$ and $\mbox{Var}(y)$ (top) and $\mathbb{P}(y=0)$ (bottom) for a \STAR process defined by $z^* \sim N(\mu, \sigma^2)$ and transformation \eqref{box-cox} with $\lambda = 1/2$ for various $(\mu, \sigma)$ pairings. The dashed gray lines corresponds to $\mathbb{E}[y]=\mbox{Var}(y)$ (top) and $\mathbb{P}(y=0) = 0$ (bottom). \STAR processes may include underdispersion (top left), overdispersion (top right), and zero-inflation (bottom). 
\label{fig:star-summary-sqrt}}
\end{center}
\end{figure}

Lastly, we show that \STAR processes are capable of modeling over- or underdispersion. In Figure~\ref{fig:star-summary-sqrt}, we illustrate the relationships among the expectation $\mathbb{E}[y]$, the variance $\mbox{Var}(y)$, and the probability of zeros $\mathbb{P}(y=0)$ for a \STAR process defined by $z^* \sim N(\mu,\sigma^2)$ and transformation \eqref{box-cox} with $\lambda = 1/2$. For different values of the parameters the \STAR process exhibits different features, including overdispersion, underdispersion, and zero-inflation.  

\subsection{Posterior inference}\label{mcmc}
We develop a general Markov Chain Monte Carlo (MCMC) algorithm for Bayesian inference under \STARp. The hierarchical construction of \STAR  in \eqref{round}-\eqref{transform} is accompanied by a computationally convenient data augmentation strategy, which we leverage to  incorporate existing sampling techniques for the unknown parameters $\theta$ in \eqref{transform}. To emphasize the modularity of the proposed approach, we omit model-specific details for sampling $\theta$ until Section~\ref{model-ex}. 

%

Let $\mathcal{D} = \{x_i, y_i\}_{i=1}^n$ denote the observed pairs of points $x_i \in \mathcal{X}$ and integer-valued data $y_i = y(x_i)$. Consider a Bayesian specification of \eqref{transform} with suitable prior on  $\theta$  and an algorithm $\mathbb{A}$ which draws from the posterior distribution of $\theta$ given the (continuous) data.  The sampling algorithm $\mathbb{A}$ is designed for continuous data, such as (Gaussian) additive models or \bartp, depending on the different choices for $\Pi_\theta$ (see Section~\ref{model-ex} for details). 
The posterior sampling algorithm for \STAR defines a Gibbs sampler by combining a data augmentation step with algorithm $\mathbb{A}$ as follows:
\begin{enumerate}
\item Sample $\left[z^*(x_i) \mid \mathcal{D}, \theta\right]$ from $\Pi_\theta$ truncated to  $z^*(x_i) \in g(\mathcal{A}_{y_i})$ for $i =1, \dots n$; 
\item Sample $\left[\theta \mid \bm z^* \right]$ using algorithm $\mathbb{A}$ conditioning on $\bm z^* = (z^*(x_1),\ldots, z^*(x_n))'$.
\end{enumerate}
In the case of model \eqref{simpleGP}, the data augmentation step may be computed efficiently using a standard univariate truncated normal sampler. 
Specifically, for $\mathcal{A}_j = [a_j, a_{j+1})$, the full conditional distribution of the latent data is $\left[z^*(x_i) \mid \mathcal{D}, \theta\right] \sim N(\mu(x_i), \sigma^2(x_i))$ truncated to $[g(a_{y_i}), g(a_{y_i + 1}))$. 
While the process $y^*$ is useful for interpretability of the \STAR model, it is not necessary for inference or sampling. 



When the transformation $g$ is unknown, an additional sampling step is required. For the parametric Box-Cox case \eqref{box-cox} this translates to sampling the parameter $\lambda$ from its full conditional posterior distribution, for which we use a slice sampler \citep{neal2003slice}. For the nonparametric model \eqref{nonparag} with prior \eqref{prior-gamma}, we sample $\xi_\gamma = \log(\tilde \gamma)$ using Metropolis-Hastings and $\big[\sigma_\gamma^{-2} \mid - \big] \sim \mbox{Gamma}\{0.001 + L/2, 0.001 + \sum_{\ell=1}^L (\tilde \gamma_{\ell} - \mu_{\gamma_\ell})^2\}$, and set $g(t) = b_I'(t)\gamma$ as defined in \eqref{nonparag}-\eqref{prior-gamma}. 
The sampler for $\xi_\gamma$ uses a Gaussian random walk proposal with covariance matrix tuned using the robust adaptive Metropolis (RAM) algorithm of \cite{vihola2012robust} during a preliminary burn-in period. Within the RAM algorithm we set a target acceptance rate of 30\% with an adaptation rate of 0.75; see  \cite{vihola2012robust} for details. We adapt the proposal covariance only during the first 50\% of the burn-in period, so the MCMC draws we save for inference are generated from a (non-adaptive) Metropolis-within-Gibbs sampling algorithm.

To simulate from the posterior predictive distribution $[\tilde y(x) | \mathcal{D}]$, we additionally sample $\left[\tilde z^*(x) \mid \theta\right]$ from $\Pi_\theta$ using the current draw of $\theta$ and set $\tilde y(x) =h\left[g^{-1}\left\{\tilde{z}^*(x)\right\}\right]$ for each $x$. This step is extremely simple, yet provides inference for integer-valued predictions, model-based imputation of missing data at  $x \in \mathcal{X}$, and useful model diagnostics.  For parametric $g$  in \eqref{box-cox}, the functions $g^{-1}(s; \lambda)$ are known, while for nonparametric $g$ we approximate $g^{-1}(s)  \approx \arg\min_t | s  -  b_I'(t) \gamma|$, where the minimum $t$ is computed over a grid of values. 



The proposed framework for MCMC balances modularity and flexibility: it combines existing algorithms for continuous data models with a transformation to provide distributional flexibility for integer-valued data. The importance of modularity has been demonstrated recently for the negative-binomial distribution, for which \cite{polson2013bayesian} developed a P{\'o}lya-Gamma data augmentation scheme for Gibbs sampling. This approach has allowed a variety of Gaussian models to be extended for negative-binomial data, including linear regression \citep{zhou2012lognormal}, factor models \citep{klami2015polya}, and functional time series models \citep{kowal2019measles}, yet faces two important limitations: first, it is restricted to the negative-binomial distribution, and second, the resulting MCMC sampler is often inefficient  \citep{duan2018scaling}. As demonstrated in Section~\ref{sec-dolphins} and Appendix~\ref{app:dolphins}, the proposed \STAR algorithm provides excellent MCMC efficiency, even for nonlinear versions of model  \eqref{simpleGP}.

\section{Regression modeling with STAR}\label{model-ex}
For inference and prediction of integer-valued data $y$ observed with predictors $x$, we apply the \STAR modeling framework to develop additive (Section~\ref{sec-am}) and \bart (Section~\ref{sec-bart}) regression models. Each model may be combined with a known or unknown transformation, and posterior inference proceeds using the general approach from Section~\ref{mcmc}. The additive and \bart \STAR models are evaluated for synthetic data in Section~\ref{sims} and real data in Sections~\ref{sec-nmes}~and~\ref{sec-dolphins}, with additional model comparisons and diagnostics in the Appendix.


%


\subsection{Additive models}\label{sec-am}
Suppose the $p$ predictors are partitioned as $x' = (u', v')$ for linear predictors $u$ and nonlinear predictors $v$. The \STAR additive model is given by \eqref{simpleGP} with conditional mean
\begin{equation}\label{am}
\mu(x) = u'\beta + \sum_j f_j(v_j),
\end{equation}
where $f_j \!:\mathcal{X}_j\to \mathbb{R}$ is an unknown function of $v_j \in \mathcal{X}_j$. The unknown $f_j$ are typically modeled as smooth nonparametric functions, and may capture nonlinearities in each $v_j$. The \STAR linear model is a special case of \eqref{am} with $\mu(x) = x'\beta$. For the conditional variance of $z^*$ in \eqref{simpleGP}, we use the conditionally conjugate prior $\sigma^{-2} \sim \mbox{Gamma}(0.001, 0.001)$.

Within the \STAR framework, we apply flexible and computationally efficient parametrizations of \eqref{am} that have been well-developed for Gaussian and  exponential family models. The linear regression coefficients are assigned conditionally Gaussian priors, $\beta \sim N(0, \Sigma_\beta)$, including the ridge prior $\Sigma_\beta = \sigma_\beta^2 I$ and other shrinkage priors \citep{carvalho2010horseshoe} as special cases. The nonlinear functions in \eqref{am} are modeled smoothly using a basis expansion $f_j(v_j) =  b_j'(v_j)  \alpha_j$, where $ b_j$ is an $L_j$-dimensional vector of known basis functions and $\alpha_j$ is a vector of unknown coefficients. We select a cubic P-spline basis with second-order difference penalty on the coefficients, which may be reparametrized such that $\alpha_j \sim N(0, \sigma_{\alpha_j}^2 I)$ is the smoothing prior and $B_j'B_j$ is diagonal, where $B_j = (b_j(v_{1,j}), \ldots, b_j(v_{n,j}))'$ is the basis matrix \citep{scheipl2012spike}. The nonlinear terms are constrained such that $\sum_{i=1}^n f_j(v_{i,j}) = 0$ for identifiability, which is enforced in the reparametrization of the basis $b_j$. We let $\sigma_{\alpha_j}^{-2} \sim \mbox{Gamma}(0.1, 0.1)$, which allows the smoothness of each $f_j$ to be learned from the data.

For observed predictors $x_i' = (u_i', v_i')$, let $U$ denote the matrix of linear predictors and let $\bm f_j =  B_j  \alpha_j$. The Gibbs sampler for the \STAR additive model iterates the following full conditional distributions:
{ \begin{enumerate}

\item Sample $[z^*(x_i) \mid -] \sim N(u_i'\beta + \sum_j f_j(v_{i,j}), \sigma^2) $ truncated to  $z^*(x_i) \in g(\mathcal{A}_{y_i})$; 

\item Sample $[\beta \mid -] \sim N\big(Q_\beta^{-1} \ell_\beta, Q_\beta^{-1}\big)$ where $Q_\beta = \sigma^{-2} U'U +   \Sigma_\beta^{-1}$ and $\ell_\beta = \sigma^{-2} U'(\bm z^* - \sum_j \bm f_j)$;

\item For each $j$, sample $[ \alpha_j \mid -] \sim N\big(Q_{\alpha_j}^{-1} \ell_{\alpha_j}, Q_{\alpha_j}^{-1}\big)$ where $Q_{\alpha_j} = \sigma^{-2} B_j'B_j +   \sigma_{\alpha_j}^{-2}I$ and $\ell_{\alpha_j} = \sigma^{-2} B_j'(\bm z^* - U\beta - 
\sum_{k \ne j} \bm f_k)$  and set $\bm f_j = B_j \alpha_j$;

\item Sample $[\sigma^{-2} \mid  - ] \sim \mbox{Gamma}\big(0.001 + n/2, 
0.001 + ||\bm z^*  -  U \beta - \sum_j \bm f_j||^2/2\big)$;

\item For each $j$, sample $[\sigma_{\alpha_j}^{-2} \mid -] \sim \mbox{Gamma}\big(0.1 + L_j/2, 0.1 + \sum_{\ell=1}^{L_j} \alpha_j^2/2\big)$.

\end{enumerate} }
The MCMC sampling algorithm for the \STAR additive model is efficient, with empirical support provided in Section~\ref{sec-dolphins}. The computational complexity for the nonlinear basis coefficients $\{\alpha_j\}$ is $\mathcal{O}(\sum_j L_j)$ due to the diagonality of $B_j'B_j$, while the linear coefficients $\beta$ also may be sampled efficiently \citep{rue2001fast,bhattacharya2016fast}.  Additional sampling steps for $\Sigma_\beta$ depend on the model specification, but are often available in closed form. 

\subsection{Bayesian Additive Regression Trees}\label{sec-bart}
While additive models are effective at capturing nonlinear marginal effects, they are often inadequate for modeling interactions among predictors. Specific pairwise or higher order interactions may be specified in advance, but including all possible interactions in an additive model requires a massive number of parameters. As a remedy, \cite{chipman2010bart} proposed \bartp, which is a ``sum-of-trees" model within a fully Bayesian framework. Tree-based regression models, such as \cite{chipman1998bayesian}, are designed to model complex interactions among predictors. Notably, \bart utilizes many trees, where each individual tree is constrained via the prior to be a weak learner. As a result, \bart provides the capability to capture nonlinear interactions yet features built-in mechanisms to guard against overfitting. For continuous and binary data, the predictive performance of \bart is highly competitive with state-of-the-art statistical and machine learning models. 

For integer-valued data, \bart has been relatively underutilized. Adaptations of \bart for negative-binomial data are feasible via P{\'o}lya-Gamma augmentation \citep{polson2013bayesian}, similar to the probit implementation in \cite{chipman2010bart} for binary data. However, this approach is limited in distributional flexibility, and the MCMC inefficiencies of P{\'o}lya-Gamma augmentation are unlikely to be ameliorated given the complexity of the (Gaussian) \bart sampling algorithm. Recently, \cite{murray2017log} proposed a log-linear \bart model for count-valued and categorical data using a parameter expansion valid for certain likelihoods in log-linear models, in particular (zero-inflated) negative-binomial and Poisson. However, extensions to more flexible count distributions may require alternative computational strategies and appropriately modified prior distributions. 

Within the \STAR framework, we parametrize the \bart model (\bartp-\STARp) as in 
\cite{chipman2010bart} and specifically
\begin{equation}\label{BART}
\mu(x) = \sum_{k=1}^m f(x; T_k, M_k),
\end{equation}
where $T_k$ is a binary tree comprised of interior splitting rules and terminal nodes and  $M_k = \{\eta_{1,k},\ldots, \eta_{b_k,k}\}$ is the value at each of $b_k$ terminal nodes for tree $T_k$. For a given predictor $x$, each tree $T_k$ in \eqref{BART} assigns a value $\eta_{\ell,k} \in M_k$, and these values are summed across all trees $k=1,\ldots,m$. \cite{chipman2010bart} propose prior distributions that constrain each $T_k$ to be shallow, thereby limiting the order of interactions, and constrain each $\eta_{\ell,k}$ to be small, thereby limiting the contribution of each tree. Both mechanisms guard against overfitting, and in combination produce a sum of weak learners. The joint prior distribution is specified as a prior for the tree, $p(T_k)$, which follows \cite{chipman1998bayesian}, and a prior for the terminal values given the tree, $p(\eta_{\ell,k} | T_k)$, which is Gaussian.  A key feature of  \STAR is that, by transforming to Gaussianity, we inherit the same framework as the original \bartp, and therefore may directly incorporate the well-studied priors and hyperparameters from \cite{chipman2010bart}. 

More careful consideration is required for the prior distribution of  $\sigma^2$ in \eqref{simpleGP}.  \cite{chipman2010bart} emphasize that an informative prior distribution is important to balance between overly aggressive and overly conservative model fits, and parametrize the prior for $\sigma^2$ as an inverse chi-square distribution calibrated using a data-based overestimate  $\hat \sigma$  of $\sigma$. However, any statistics calculated from the data $y$ are likely inappropriate for \STARp, since the transformation $g$ impacts the scale of $z^*$. As a remedy, we compute $\hat \sigma$ as the posterior median of $\sigma$ from the \STAR linear model \eqref{am}, where the transformation in the linear model is chosen to match the transformation in \bartp-\STARp. Given $\hat \sigma$, which indeed is a data-based overestimate of $\sigma$, we adopt the default hyperparameter suggestions of \cite{chipman2010bart}.

For posterior inference under \bartp-\STARp, we combine a data augmentation step for $z^*(x_i)$ similar to  Section~\ref{sec-am} with a sweep from the original \cite{chipman2010bart} algorithm to draw the \bart parameters in \eqref{BART} using $\bm z^*$ as data. The \cite{chipman2010bart} \bart sampler proceeds using backfitting: draws for the $k$th tree $[(T_k, M_k) |  \bm z^*, \{(T_{k'}, M_{k'})\}_{k \ne k'}, \sigma]$ are generated using \cite{chipman1998bayesian} and $\sigma^2$ is sampled from an inverse-Gamma distribution. Incorporating these sampling steps  into the larger Gibbs sampler in Section~\ref{mcmc} is straightforward using the \texttt{dbarts} package \citep{dbarts}.

\section{Simulation studies}\label{sims}
The proposed \STAR modeling framework is evaluated using simulated data and compared to existing methods for  Poisson, negative-binomial, and Gaussian data. Synthetic data $y_i$ for $i = 1, \dots n$ and $n=100$ are simulated from a negative-binomial distribution parametrized by conditional mean $\mathbb{E}\{y_i(x) | \lambda_i^*\} = \lambda_i^*(x)$ and variance $\mbox{Var}\{y_i(x) | \lambda_i^*, r^*\} = \lambda_i^*(x)\left\{1 + \lambda_i^*(x)/r^*\right\}$ with dispersion parameter $r^* > 0$. 
As $r^*$ decreases to zero, the variance increasingly dominates the mean while as $r^*\rightarrow \infty$, the distribution converges to a Poisson distribution with parameter $\lambda_i^*(x)$. We select $r^*=1$ to simulate negative-binomial data with large overdispersion and $r^*=1000$ to simulate approximate Poisson data. We consider linear (Section~\ref{sim:lm}) and nonlinear (Section~\ref{sim:bart}) parametrizations for the log-mean $\log \lambda_i^*(x)$.
We emphasize that in all cases, the simulated datasets are \emph{not} generated under the proposed \STAR model: they are simulated from negative-binomial and (approximate) Poisson distributions.

Competing models are compared using the Watanabe-Akaike/widely-applicable information criteria (WAIC)  \citep{watanabe2010asymptotic}. WAIC estimates out-of-sample predictive accuracy using a single model fit requiring only minimal additional computations, and is asymptotically equivalent to cross-validation. The WAIC for a model $\mathcal M$ is defined as $
\mbox{WAIC}_{\mathcal M} = -2\left(\mbox{lpd}_{\mathcal M} - d_{\mathcal M} \right)$, 
where $d_{\mathcal M}$ is the effective number of parameters for model ${\mathcal M}$ and $\mbox{lpd}_{\mathcal M}$ is the log-predictive pointwise density defined by 
\[
\mbox{lpd}_{\mathcal M}(y) = \sum_{i=1}^n \log \left( \frac{1}{S} \sum_{s=1}^S p_{\mathcal M}(y_i \mid \theta^s) \right)
\]
for $\theta^s$ drawn from its posterior distribution. For \STAR models, we simply have
\begin{equation}\label{star-lppd}
\mbox{lpd}_{\STARp}(y) = \sum_{i=1}^n \log \left( \frac{1}{S} \sum_{s=1}^S  
\Phi\left\{ \frac{g^s(a_{y_i+1}) - \mu^s(x_i)}{\sigma^s(x_i)}\right\} - 
\Phi\left\{ \frac{g^s(a_{y_i}) - \mu^s(x_i)}{\sigma^s(x_i)}\right\} 
\right).
\end{equation}
For the effective number of parameters, we follow the recommendation of \cite{gelman2014understanding} and use the sample variance of the pointwise log-likelihoods across MCMC simulations: 
$d_{\mathcal M}  = \sum_{i=1}^n \mbox{Var}\left( \log p(y_i \mid \theta^s)\right)$.
The pointwise log-likelihood of \STAR is simple and efficient to compute, and is sufficient for computing WAIC as well as other information criteria.

WAIC is used in Sections~\ref{sim:lm}~and~\ref{sim:bart} for synthetic data, but exact out-of-sample metrics are provided for the real data in Section~\ref{sec-nmes}. Appendix~\ref{app:sims} also provides root mean square errors for estimating the conditional expectation $\lambda_i^*(x)$ to evaluate point estimation accuracy. 

\subsection{Linear mean functions}\label{sim:lm}
We first consider a linear log-mean,  $\log \lambda_i^*(x) = \beta_0 + \sum_{j=1}^p x_{i,j} \beta_j$, where the $p=6$ predictors are drawn independently from $x_{i,j} \sim N(0,1)$ and the coefficients are  $\beta_0 = \log(1.5)$, $\beta_1 = \beta_2 =\beta_3 = \log(2.0)$, and   $\beta_4=\beta_5=\beta_6 = 0$. Under this specification, the expected counts at $x_{i,j} = 0$ is 1.5, while each nonzero coefficient $\beta_j$ for $j=1,2,3$ increases the expected counts by a factor of 2 per one unit change in each $x_j$.

For comparison, we consider a variety of Bayesian linear regression models.  Among \STAR models, we use the linear model \eqref{am} with prior $[\beta_j | \sigma_\beta] \sim N(0, \sigma_\beta^2)$ and $[\sigma_\beta] \sim \mbox{Uniform}(0, 10^4)$, and the following transformations: known transformation \eqref{box-cox} with $\lambda = 0$ (\textsc{lm-star}-log), $\lambda = 1/2$ (\textsc{lm-star}-sqrt), and $\lambda = 1$ (\textsc{lm-star}-id); unknown parametric transformation \eqref{box-cox} (\textsc{lm-star}-bc); and unknown nonparametric transformation \eqref{nonparag} (\textsc{lm-star}-np). We also include the same linear model, but  instead with the Gaussian model \eqref{simpleGP} applied directly to the counts $y$ (\textsc{lm}) and the log-transformed counts $\log(y + 1)$  (\textsc{lm}-log). These models are natural competitors to \STARp, since they incorporate the same model for $\mu(x)$ but omit the rounding step in \eqref{round} and therefore do not produce an integer-valued distribution. Lastly, we include Poisson (\textsc{lm}-Pois) and negative-binomial (\textsc{lm}-NegBin) linear regression models with a log-link, implemented using the \texttt{rstanarm} package \citep{rstanarm}. \textsc{lm}-Pois and \textsc{lm}-NegBin are widely used for modeling count data, and here correspond to the true data-generating process. 

\begin{figure}[h]
\begin{center}
\includegraphics[width=0.49\textwidth]{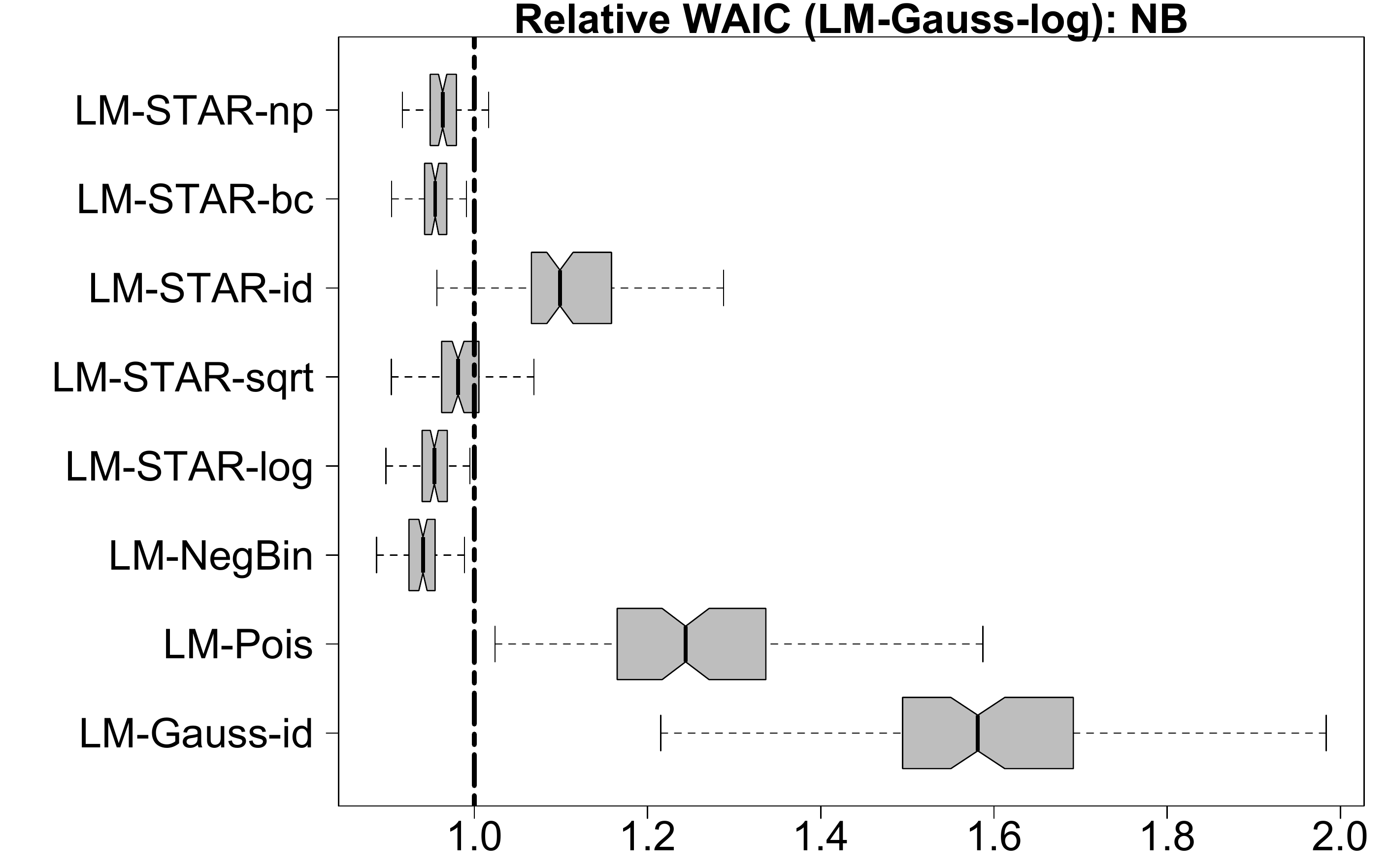}
\includegraphics[width=0.49\textwidth]{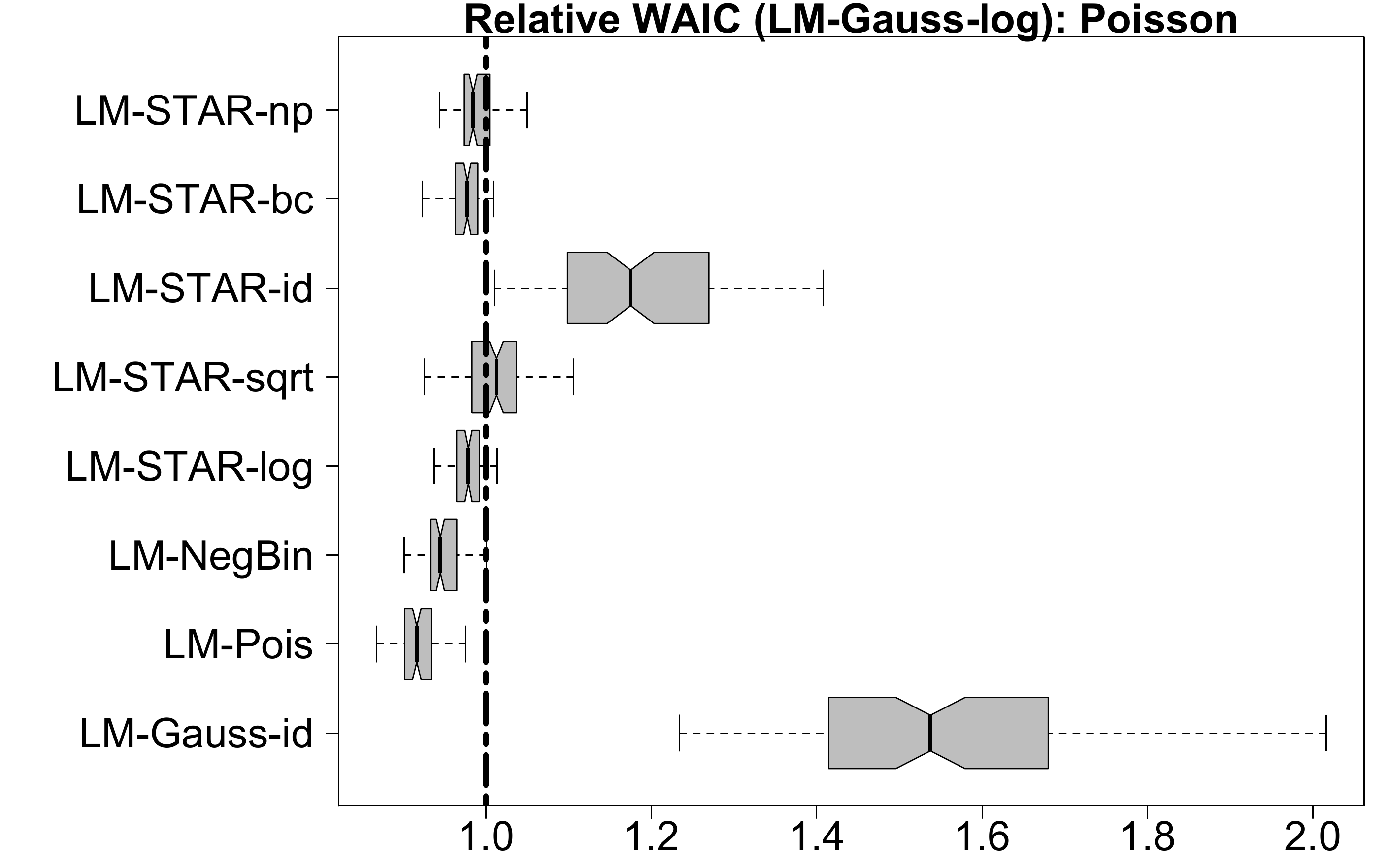}
\caption{\small Relative WAIC for negative-binomial (left) and Poisson (right) data with linear mean functions. Preferred models have smaller values, and models with values less than one are preferred to \textsc{lm}-log. The \STAR models outperform the Gaussian models and are highly competitive with the true models \textsc{lm}-Pois and \textsc{lm}-NegBin. 
\label{fig:sim-lm-waic}}
\end{center}
\end{figure}

In Figure~\ref{fig:sim-lm-waic}, we plot the relative WAICs across simulated datasets, defined as the ratio between the WAIC of the generic model over the WAIC for a baseline method, for which we select \textsc{lm}-log. 
Relative WAIC standardizes model performance across simulated datasets: methods with a relative WAIC less than one demonstrate improvement relative to the baseline method. The \STAR models, particularly those with unknown transformations (\textsc{lm-star}-bc and \textsc{lm-star}-np), offer substantial improvements relative to \textsc{lm}-log, and are highly competitive with the true models \textsc{lm}-Pois and \textsc{lm}-NegBin. The \STAR model improvements relative to  the Gaussian models and the identity transformation \textsc{lm-star}-id definitively demonstrate the importance of both rounding and transformation.

\subsection{Nonlinear mean functions}\label{sim:bart}
To evaluate \textsc{bart-star}, we specify a nonlinear form for the log-mean,   $\log \lambda_i^*(x) = \beta_0 + \beta_1 \tilde f(x)$, where  $\tilde f(x)$ is the centered and scaled Friedman function \citep{friedman1991multivariate}
\begin{equation}\label{friedman}
f(x) = 10 \sin(\pi x_1 x_2) + 20(x_3 - 0.5)^2 + 10x_4 + 5x_5
\end{equation}
featured in the original \bart simulations  \citep{chipman2010bart}. As in \cite{chipman2010bart}, we select $p=10$ and simulate $x_{i,j} \stackrel{iid}{\sim}\mbox{Uniform}(0,1)$. We fix the parameters $\beta_0 = \log(1.5)$ and $\beta_1 =\log(5.0)$, which again corresponds to low counts with a moderate signal.

We combine the \bartp-\STAR model of Section~\ref{sec-bart} with known transformation \eqref{box-cox} for $\lambda = 0$ (\textsc{bart-star}-log), $\lambda = 1/2$ (\textsc{bart-star}-sqrt), and $\lambda = 1$ (\textsc{bart-star}-id); unknown parametric transformation \eqref{box-cox} (\textsc{bart-star}-bc); and unknown nonparametric transformation \eqref{nonparag} (\textsc{bart-star}-np). For competitors, we include the Gaussian \bart model (\bartp-id) of \cite{chipman2010bart} and a Gaussian \bart model on the log-transformed counts $\log(y+1)$ (\bartp-log). Lastly, we include the linear models \textsc{lm-star}-bc and \textsc{lm}-log from Section~\ref{sim:lm}.

\begin{figure}[h]
\begin{center}
\includegraphics[width=0.49\textwidth]{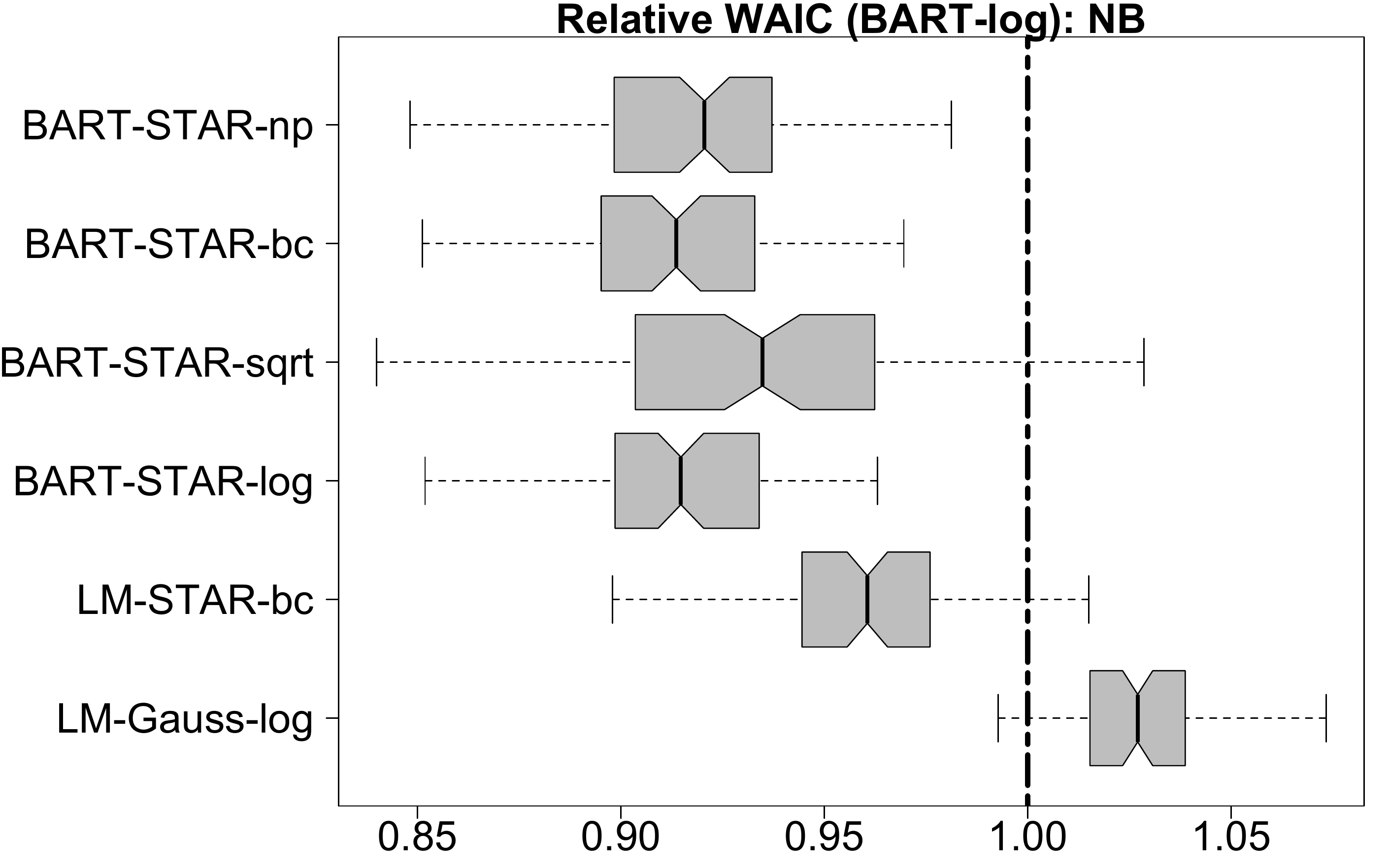}
\includegraphics[width=0.49\textwidth]{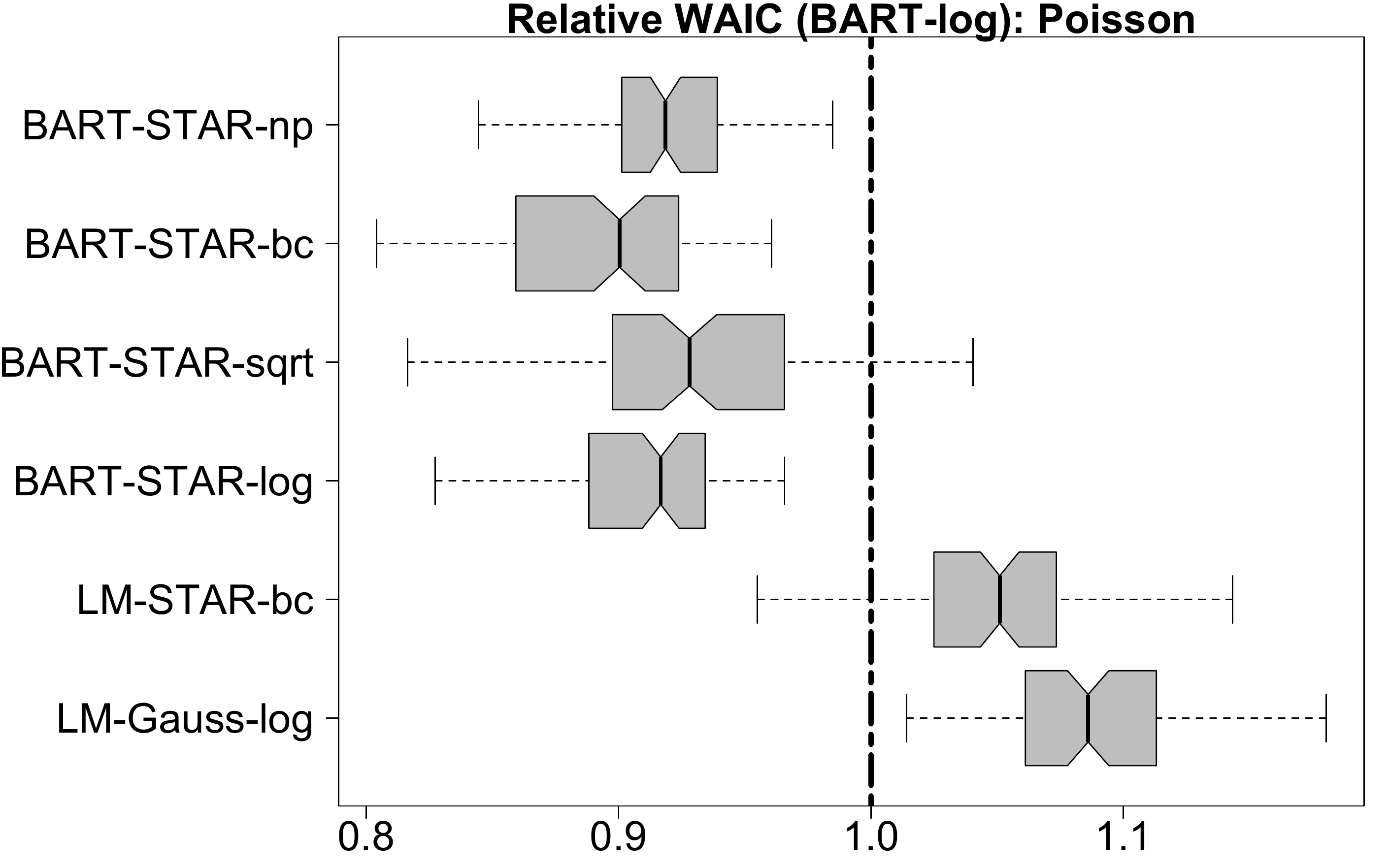}
\caption{\small 
Relative WAIC for negative-binomial (left) and Poisson (right) data with nonlinear mean functions. Preferred models have smaller values, and models with values less than one are preferred to \bartp-log. The identity models (\bartp-id, \bartp-\STARp-id) are omitted since they are noncompetitive, with relative WAICs above 1.6. The \bartp-\STAR models are clearly superior.
\label{fig:sim-bart-waic}}
\end{center}
\end{figure}

The relative WAICs are plotted in Figure~\ref{fig:sim-bart-waic}, where again we use the log-transformed Gaussian model (\bartp-log) as the baseline. Both \textsc{bart-star}-id  and \textsc{bart}-id are omitted as  noncompetitive, and indicates the importance of an appropriate transformation.  \bartp-\STAR provides substantial improvements relative to \bart and linear Gaussian models, with especially strong performance from the unknown transformation models (\textsc{bart-star}-bc and  \textsc{bart-star}-np). Perhaps surprisingly, the \STAR linear model \textsc{lm-star}-bc outperforms \bartp-log for negative-binomial data, despite the nonlinearity in \eqref{friedman}. By comparison, \textsc{bart-star}-bc consistently outperforms \textsc{lm-star}-bc, which suggests that the proposed \bartp-\STAR model is capable of detecting the nonlinear features in \eqref{friedman}.

\section{Predicting the demand for healthcare utilization}\label{sec-nmes}
Individualized prediction of healthcare utilization is critical both for assessing the health risks of an individual and for monitoring the aggregate stress on the healthcare system. By providing more accurate \emph{predictive distributions} of individual healthcare utilization, it is possible to obtain uncertainty quantification for various measures of individual and aggregate demand, and consequently achieve more efficient allocation of medical resources and more informed patients. To assess the predictive ability of \STAR models for this task, we use data from the National Medical Expenditure Survey (NMES)  available in the \texttt{AER} package in \texttt{R} \citep{aer}. Multiple measures of healthcare utilization are available, including physician office visits (\texttt{visits}), non-physician office visits (\texttt{nvisits}), physician hospital outpatient visits (\texttt{ovisits}), and non-physician hospital outpatient visits (\texttt{novisits}). Individualized predictors are also provided, including health measures, socioeconomic and demographic variables, and indicators of each patient's type of insurance. We consider a subset of $n = 4406$  elderly adults (aged 66 and older) covered by Medicare, which was previously analyzed by  \cite{deb1997demand} and  \cite{cameron2013regression}.



\begin{figure}[ht]
\begin{center}
\includegraphics[width=1\textwidth]{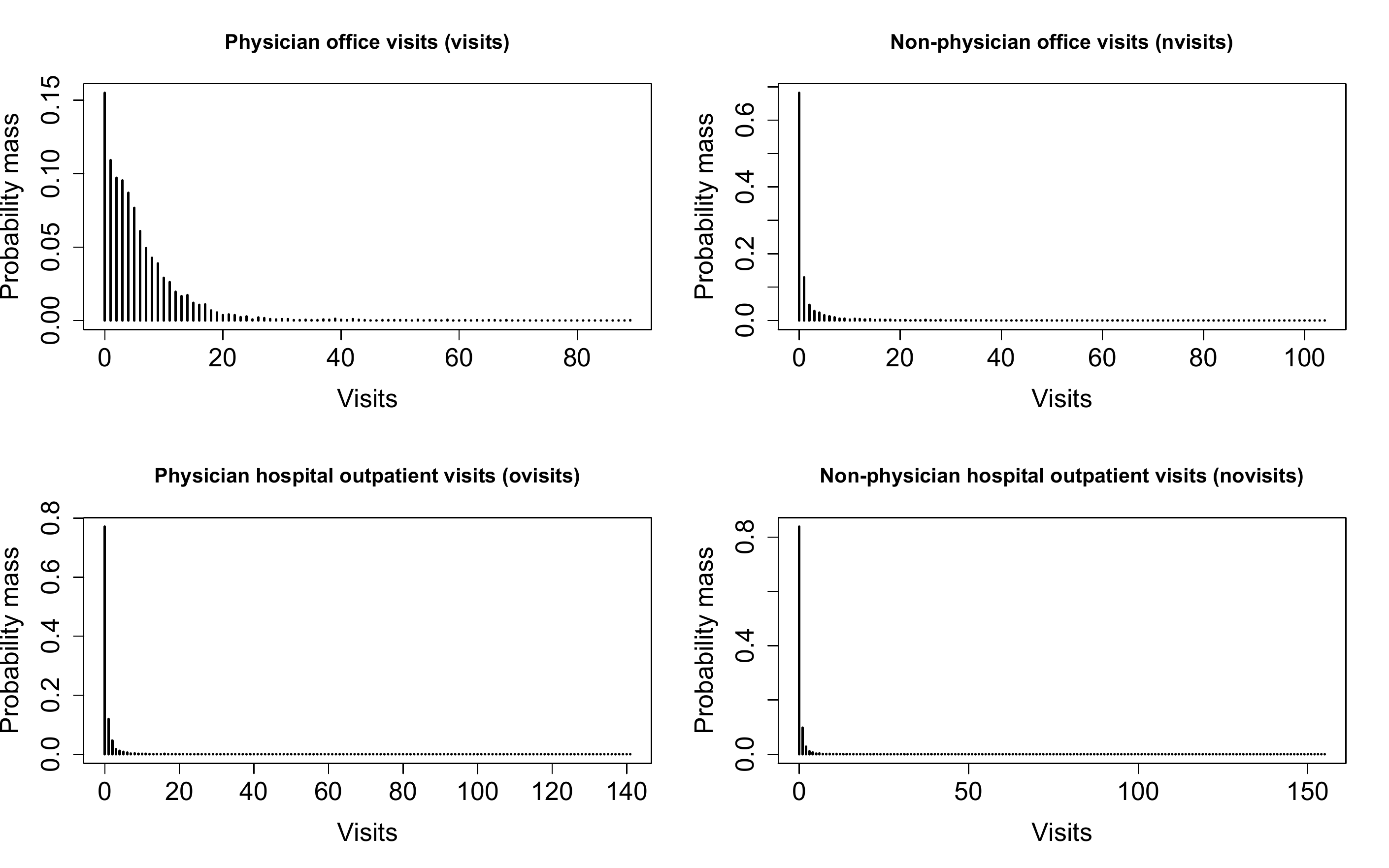}
\caption{\small Probability mass functions for each measure of medical care demand. Zero-inflation and overdispersion are present in each case to varying degrees. \label{fig:nmes}}
\end{center}
\end{figure}

The NMES data provides a unique opportunity for insightful out-of-sample prediction comparisons. Each measure of healthcare utilization  (\texttt{visits}, \texttt{nvisits}, \texttt{ovisits}, and \texttt{novisits}) is count-valued with distinct characteristics: the probability mass functions in Figure~\ref{fig:nmes} illustrate the differences in the marginal distributions, most notably the proportion of zeros and the degree of overdispersion. An adequate prediction of individual healthcare utilization may require prediction of one or more of these response variables, each of which presents unique count-valued distributional features, and which share a common set of $p=17$ individual predictor variables. 

We consider out-of-sample predictive distribution accuracy for each response variable in Figure~\ref{fig:nmes}. In all cases, we select $n_{train} = 3525$ (80\%) individuals randomly for training and evaluate the predictive accuracy for the remaining $n_{test} = 881$ test individuals, and repeat this exercise for 100 iterations. Posterior predictive distributions were computed for each model in Section~\ref{sim:bart}; for conciseness, we report results for \bartp-log, \bartp-\STARp-id, \bartp-\STARp-log, and \bartp-\STARp-bc. The (untransformed) \bart model of \cite{chipman2010bart} was noncompetitive and is omitted from the subsequent results.

\begin{figure}[h]
\begin{center}
\includegraphics[width=0.49\textwidth]{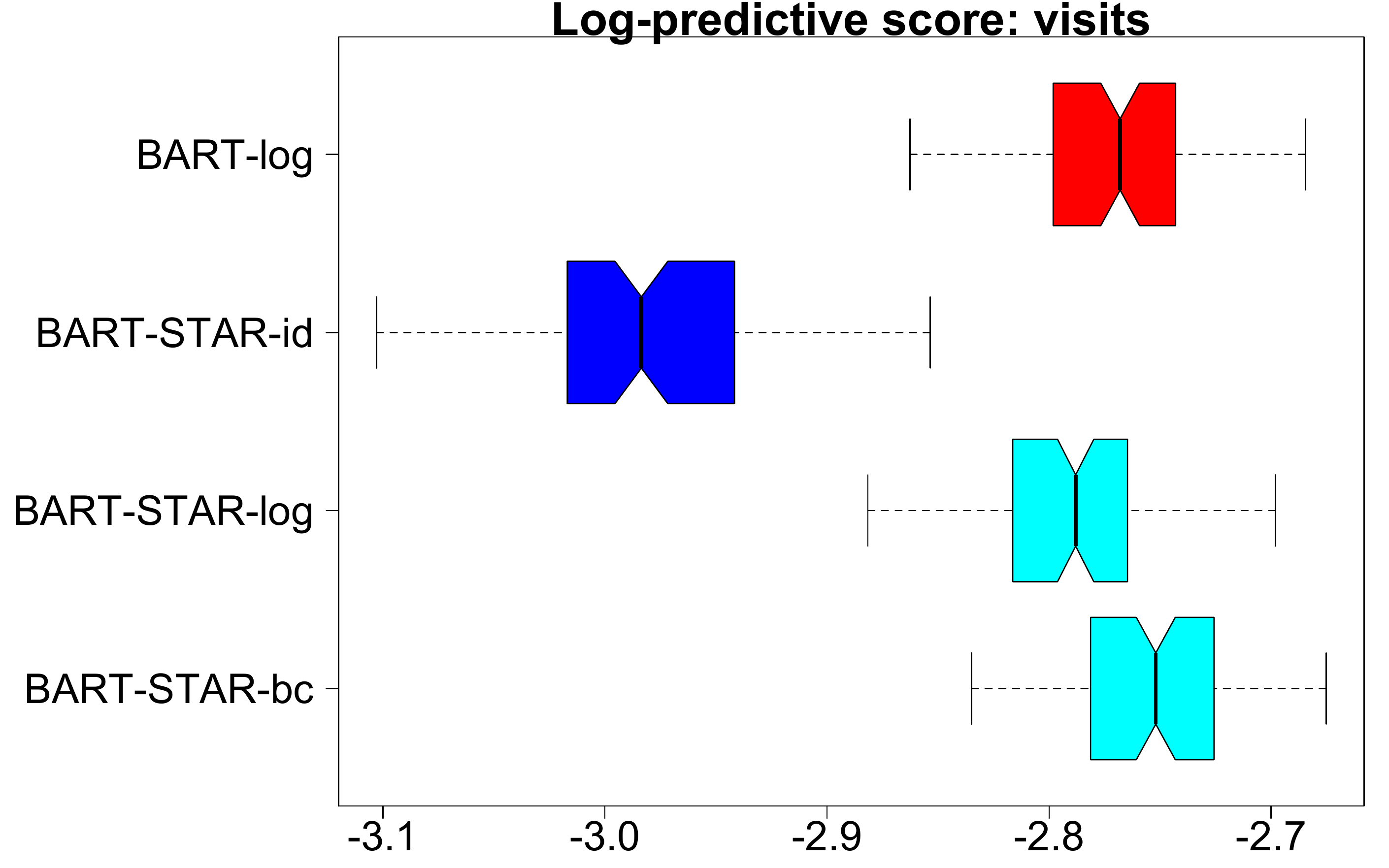}
\includegraphics[width=0.49\textwidth]{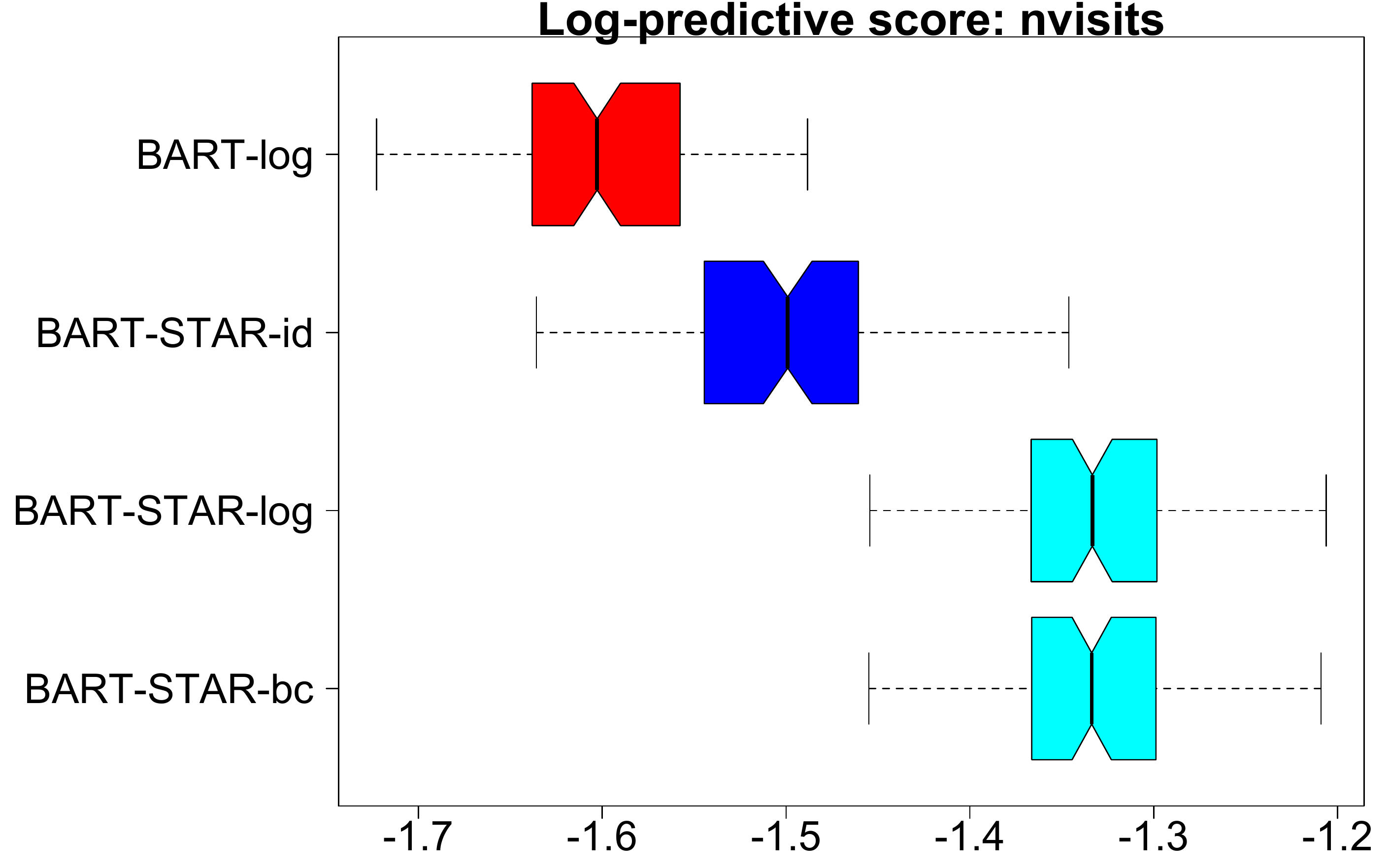}
\includegraphics[width=0.49\textwidth]{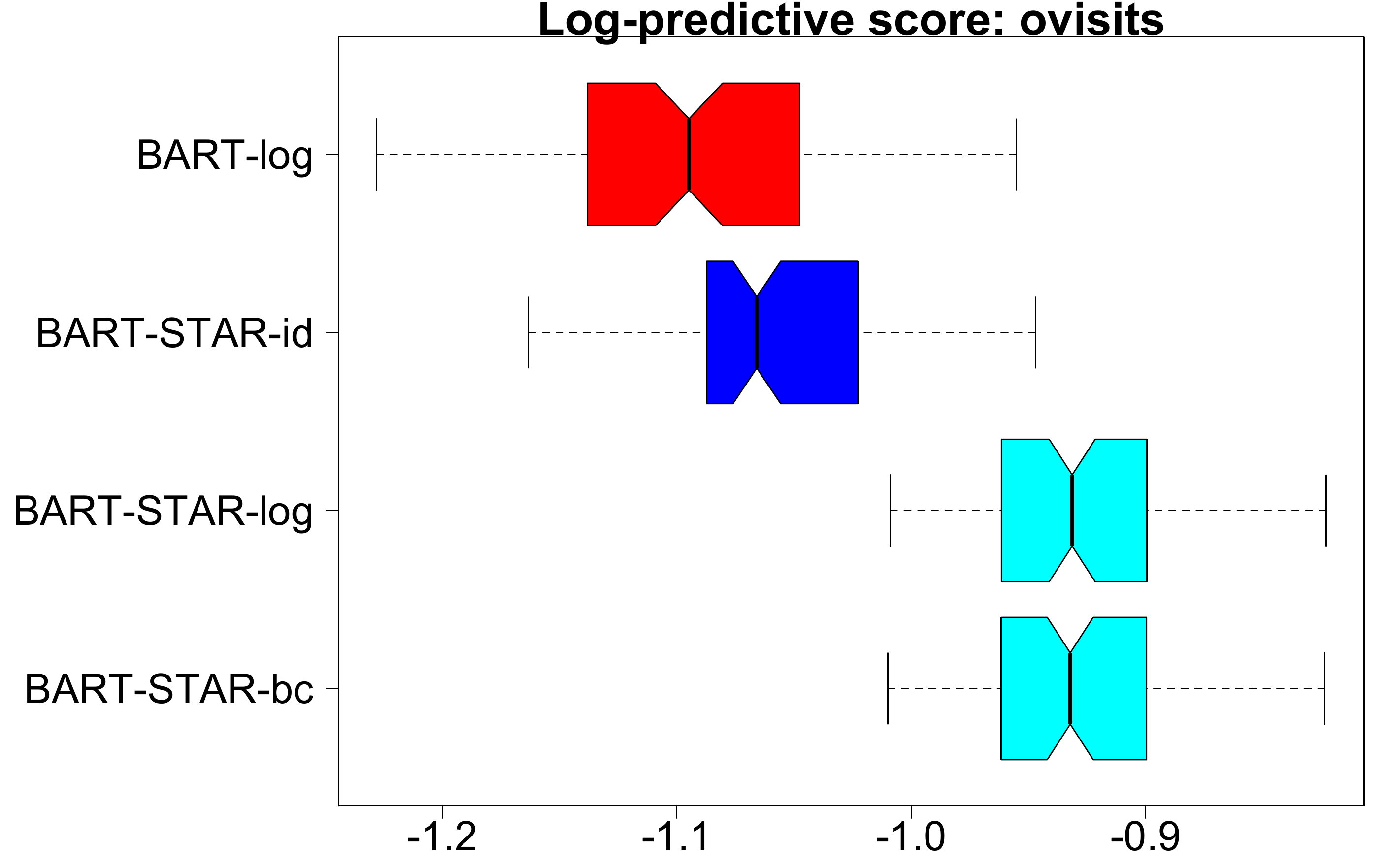}
\includegraphics[width=0.49\textwidth]{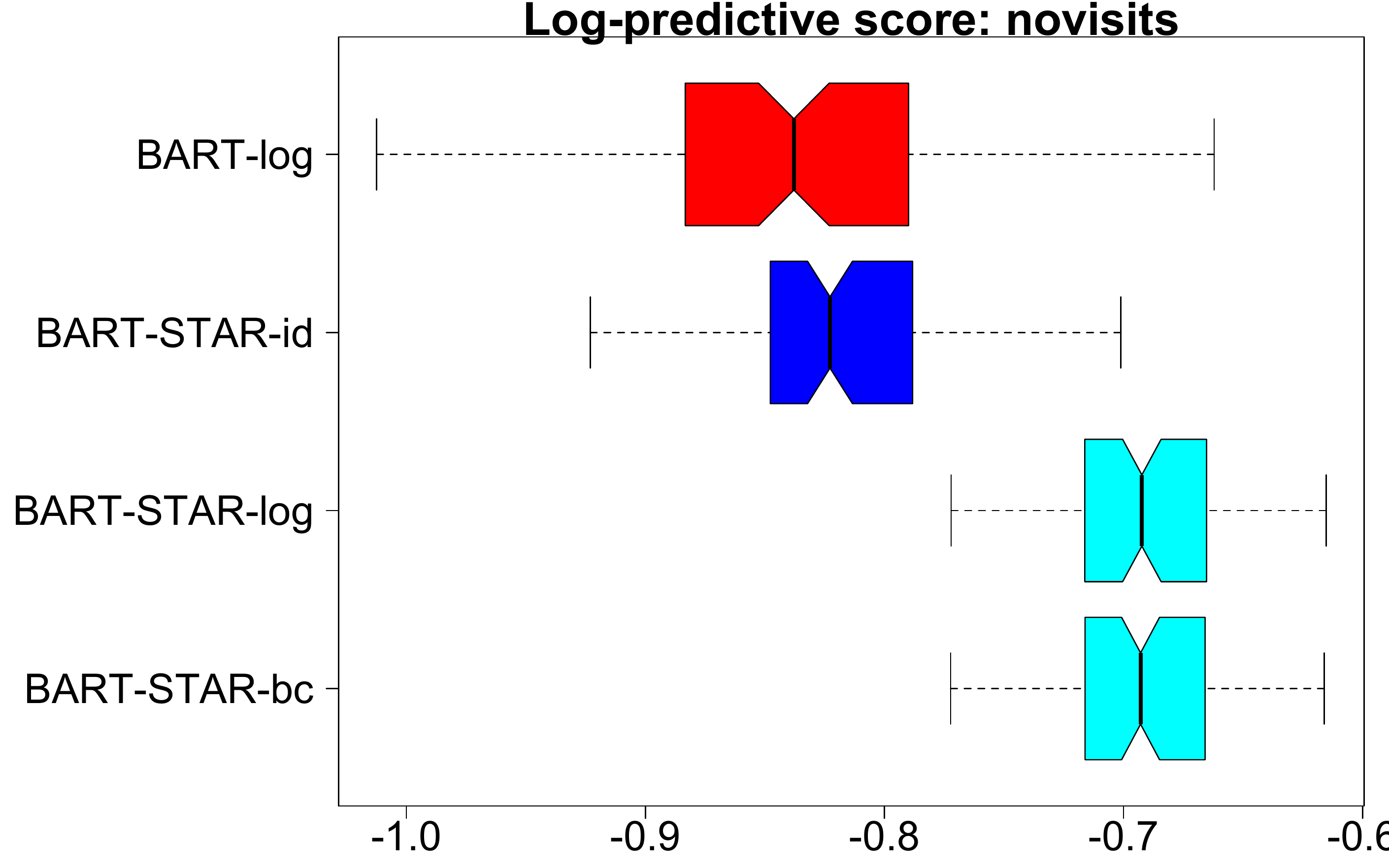}
\caption{\small Log-predictive density (LPD) score for the test data across 100 randomly selected test sets (20\% of data). Large LPD score indicates better performance. The \STAR models that include transformations---known (\STARp-log) or unknown (\STARp-bc)---are decisively favored.  
\label{fig:NMES-lpd}}
\end{center}
\end{figure}

To evaluate the aggregate predictive distribution accuracy, we compute the log-predictive density score for the test data $\{ \tilde y_i\}_{i=1}^{n_{test}}$ for each model $\mathcal{M}$:
\begin{equation}\label{lpd}
\mbox{lpd}_\mathcal{M}(\tilde y) = \frac{1}{n_{test}}\sum_{i=1}^{n_{test}} \log p_\mathcal{M}(\tilde y_i | y) \approx \frac{1}{n_{test}}\sum_{i=1}^{n_{test}} \log \left\{ \frac{1}{S} \sum_{s=1}^S p_{\mathcal M}(\tilde y_i \mid \theta^s) \right\}
\end{equation}
where $\theta^s \sim p_\mathcal{M}(\theta | y)$ is a draw from the posterior under model $\mathcal{M}$. For \STAR with model \eqref{simpleGP}, we have 
$p_{\mathcal M}(\tilde y_i \mid \theta^s) = \Phi\left[ \{g^s(a_{\tilde y_i  +1}) - \mu^s(x_i)]\}/\sigma^s(x_i)\right] -  \Phi\left[ \{g^s(a_{\tilde y_i}) - \mu^s(x_i)\}/\sigma^s(x_i)\right]$ similar to \eqref{star-lppd}. 
The results for each response variable are in Figure~\ref{fig:NMES-lpd}; larger values indicate more accurate predictive distributions. The \STAR model with unknown transformation (\bartp-\STARp-bc) performs best in all cases, closely followed by the \STAR model with known transformation (\bartp-\STARp-log). Omitting the transformation (\bartp-\STARp-id), similar to \cite{canale2013nonparametric}, leads to substantial deterioration in predictive accuracy. Furthermore, the untransformed \bartp-\STARp-id model produced infinite scores for some test points $\tilde y_i$; these points were excluded for computing \eqref{lpd}, but imply that \bartp-\STARp-id performs even worse than indicated. 

\begin{figure}[h]
\begin{center}
\includegraphics[width=0.49\textwidth]{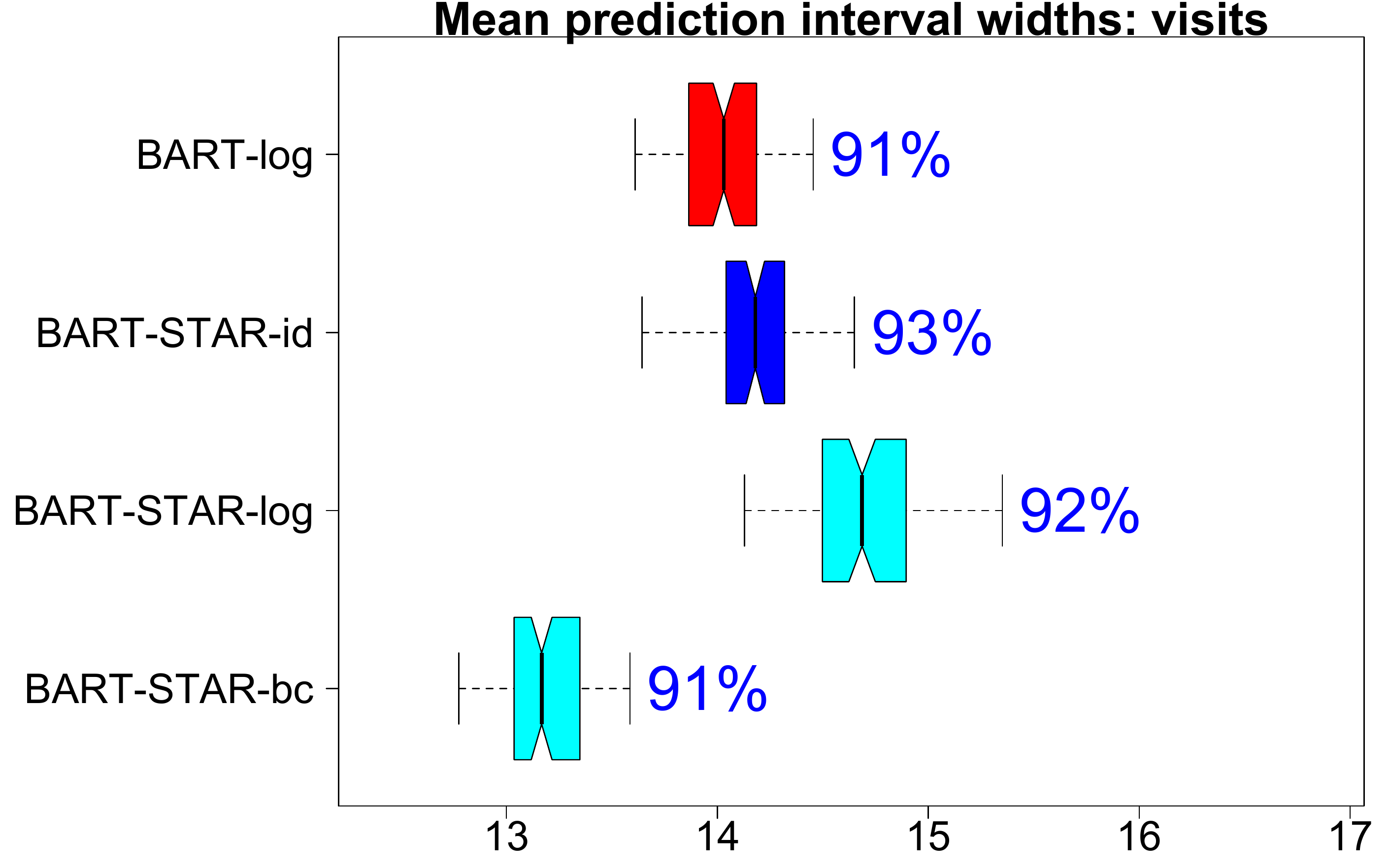}
\includegraphics[width=0.49\textwidth]{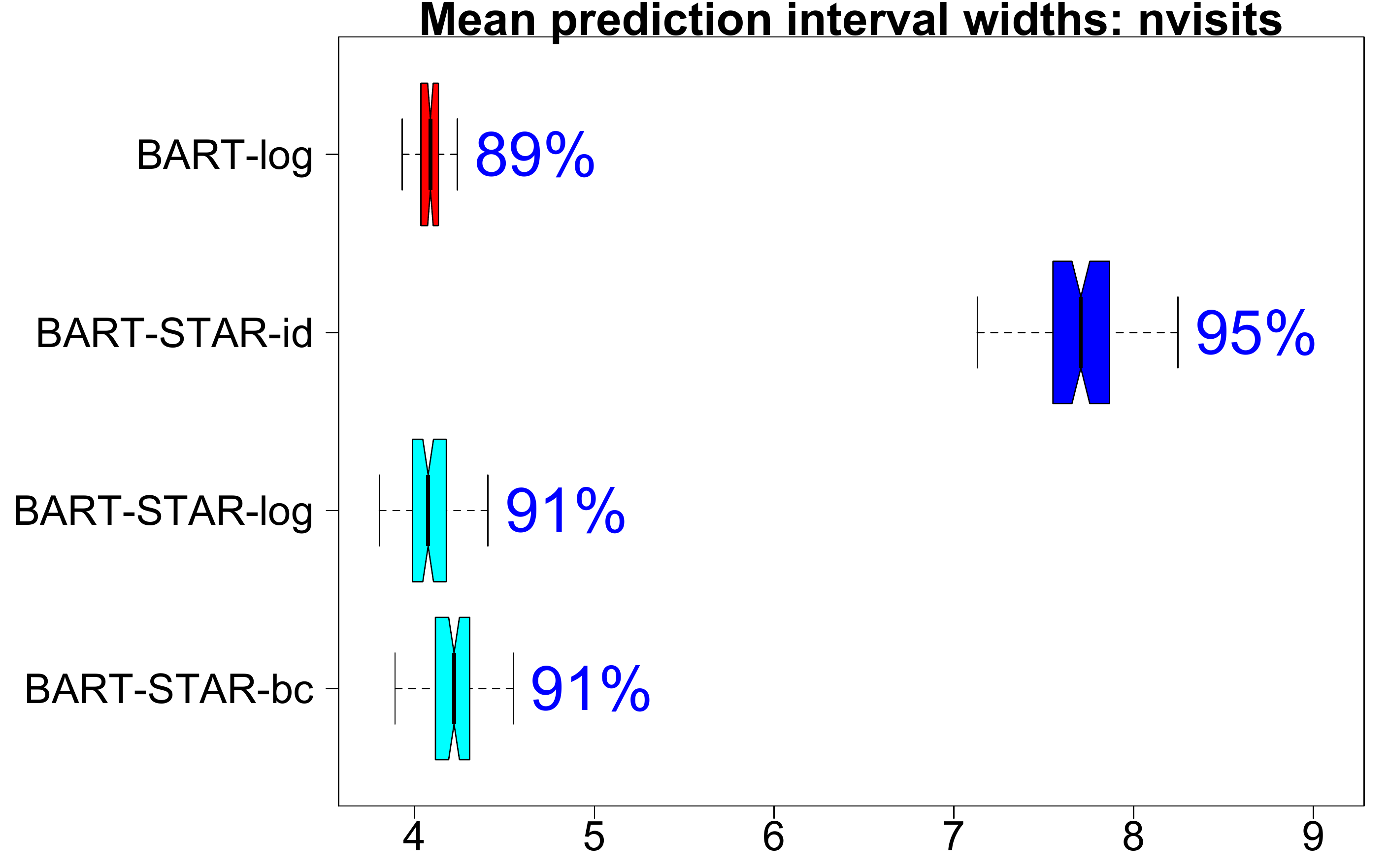}
\includegraphics[width=0.49\textwidth]{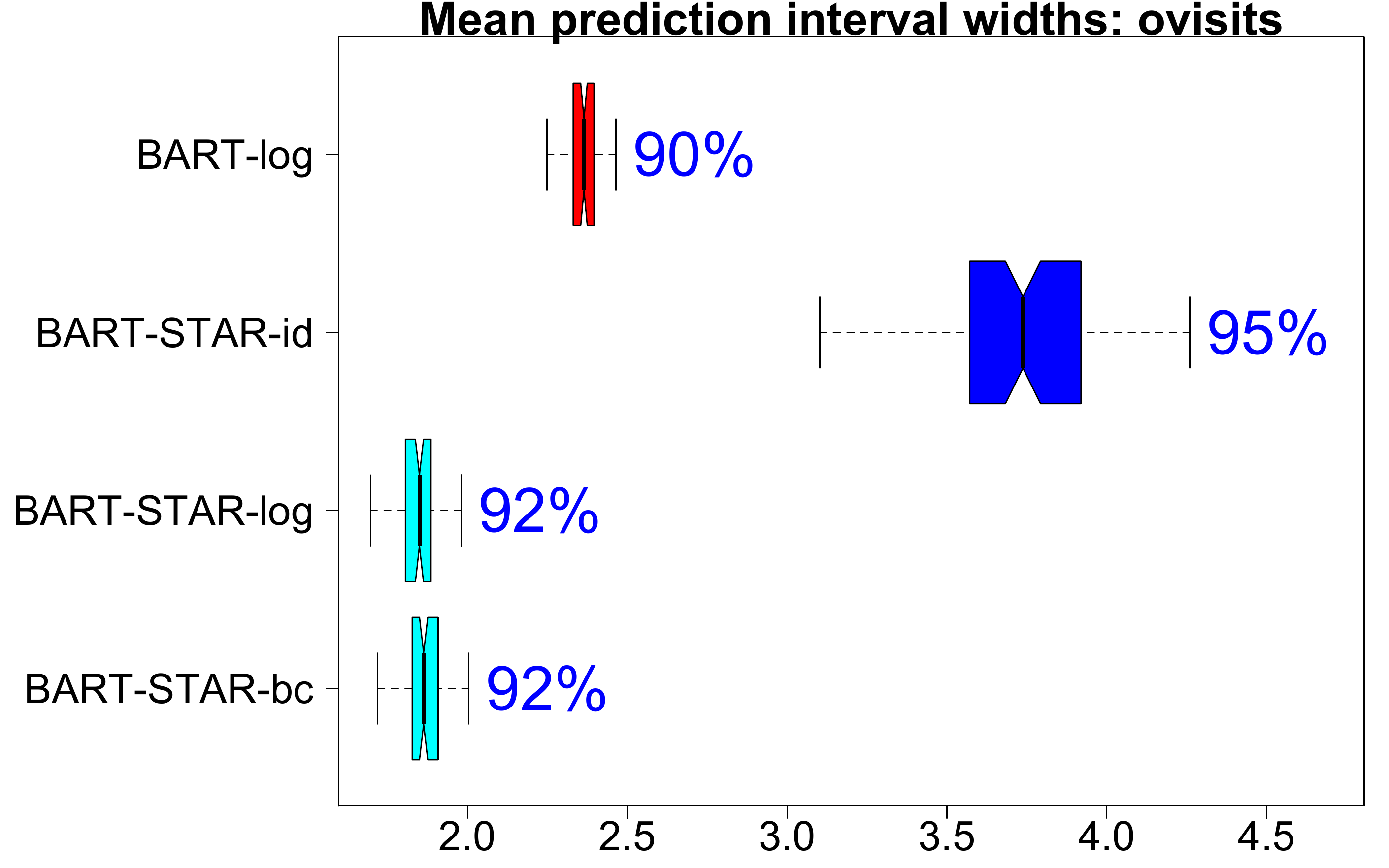}
\includegraphics[width=0.49\textwidth]{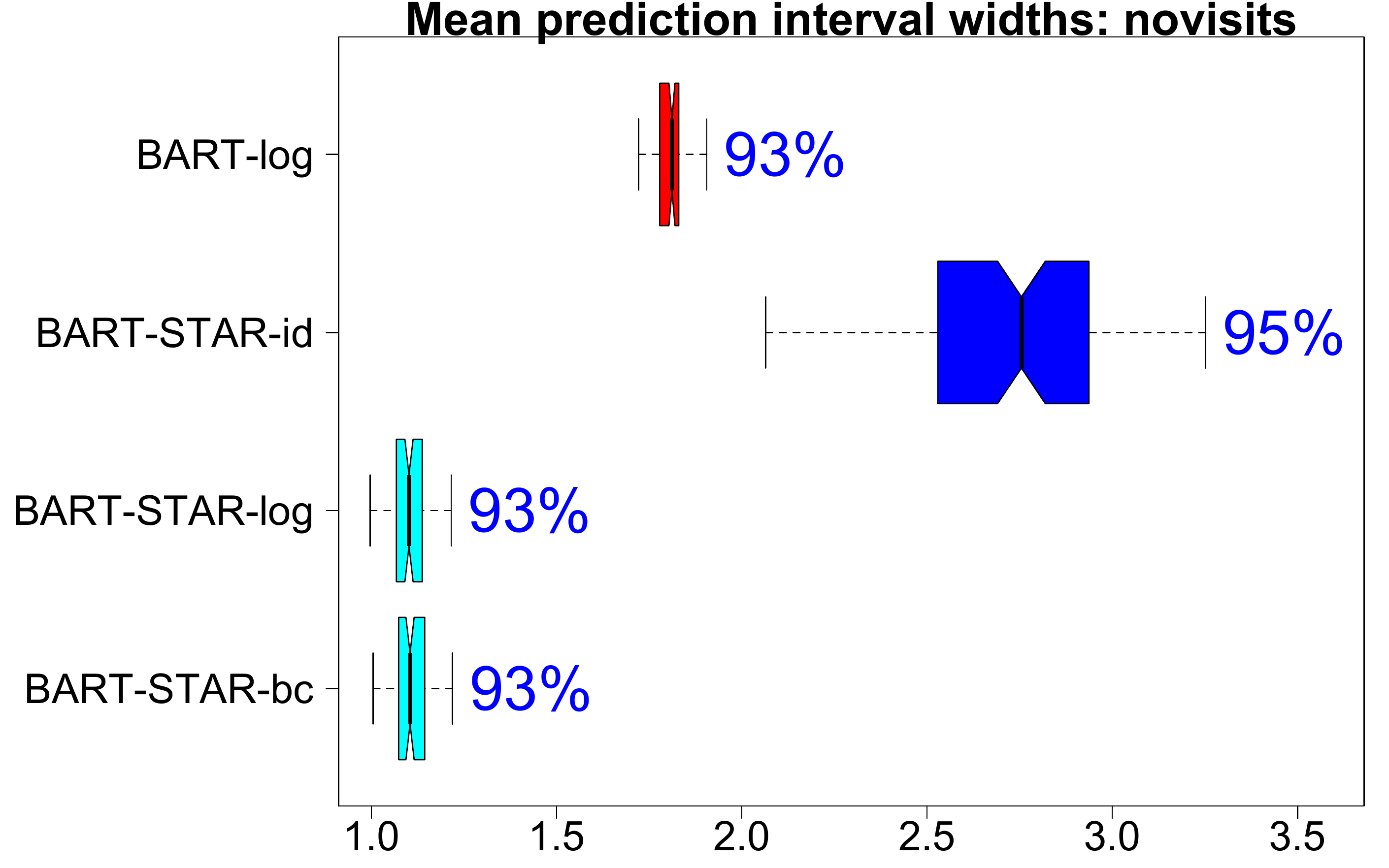}
\caption{\small Mean prediction interval widths and empirical coverage for the test data across 100  test sets. Preferred methods provide the narrowest intervals while maintaining the nominal (90\%) coverage.  The \bartp-\STAR model with unknown transformation (\bartp-\STARp-bc) is consistently competitive, followed by the \bartp-\STAR model with known log transformation (\bartp-\STARp-log). The methods that do not including rounding (\bartp-log) or include rounding \emph{without} transformation (\bartp-\STARp-id; \citealp{canale2013nonparametric}) are decisively inferior. 
\label{fig:NMES-mpiw}}
\end{center}
\end{figure}

For a more targeted assessment of the predictive distributions, we compare the precision and coverage of the 90\% prediction intervals for each method. Interval precision is measured by the mean prediction interval width (MPIW) computed across all $\{\tilde y_i\}_{i=1}^{n_{test}}$: smaller intervals that provide the correct coverage are preferable. The MPIWs with empirical coverages are displayed in Figure~\ref{fig:NMES-mpiw}. Across all responses, the \STAR model with unknown transformation (\bartp-\STARp-bc) provides the most precise prediction intervals with  correct coverage. For the responses with greater zero-inflation and overdispersion (\texttt{nvisits}, \texttt{ovisits}, and \texttt{novisits}), \bartp-\STARp-bc reduces MPIWs by a median of 45\%, 50\%, and 59\%, respectively, relative to \bartp-\STARp-id, which indicates a substantial gain in predictive precision offered by the (unknown) transformation.

\begin{figure}[h]
\begin{center}
\includegraphics[width=0.49\textwidth]{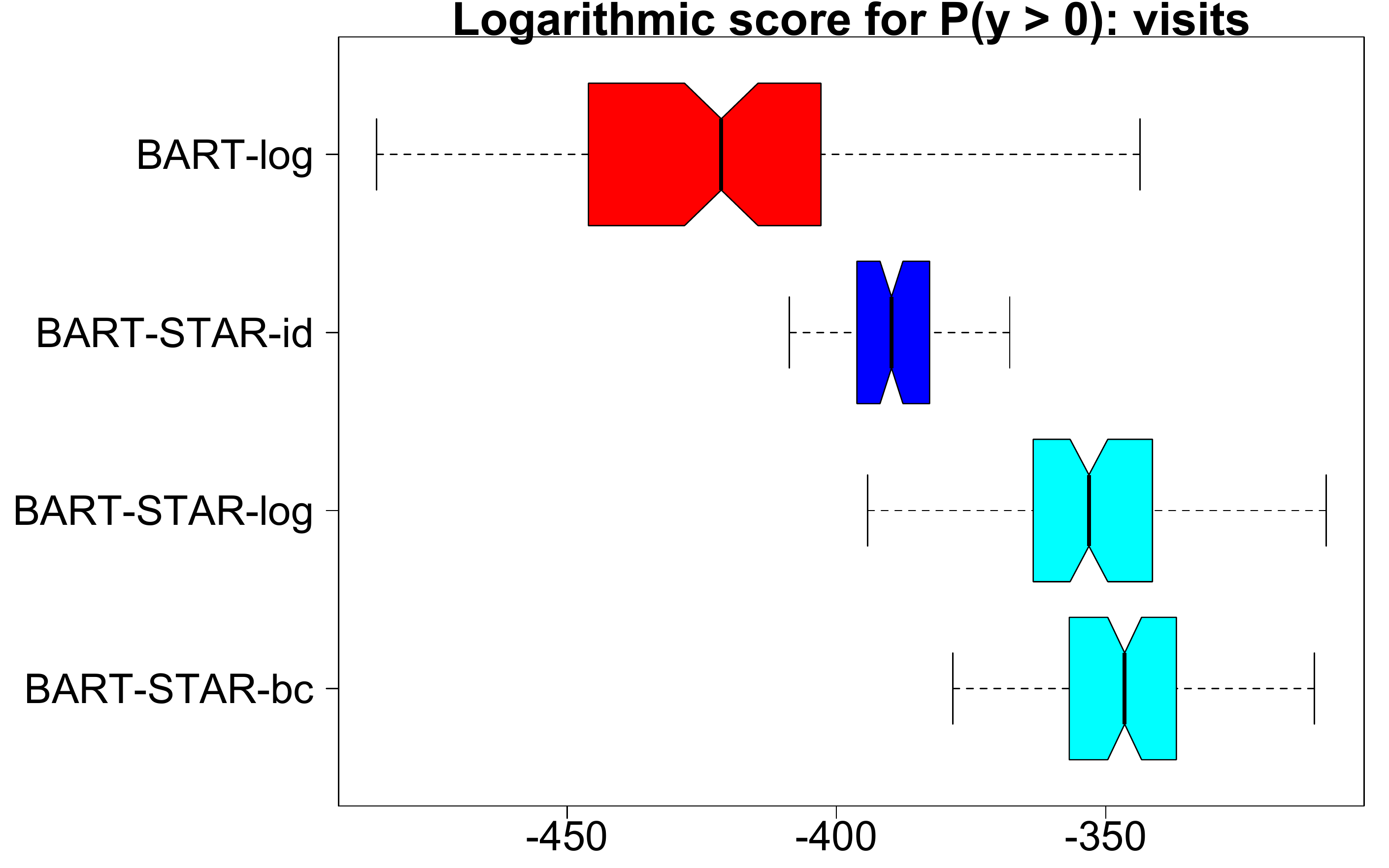}
\includegraphics[width=0.49\textwidth]{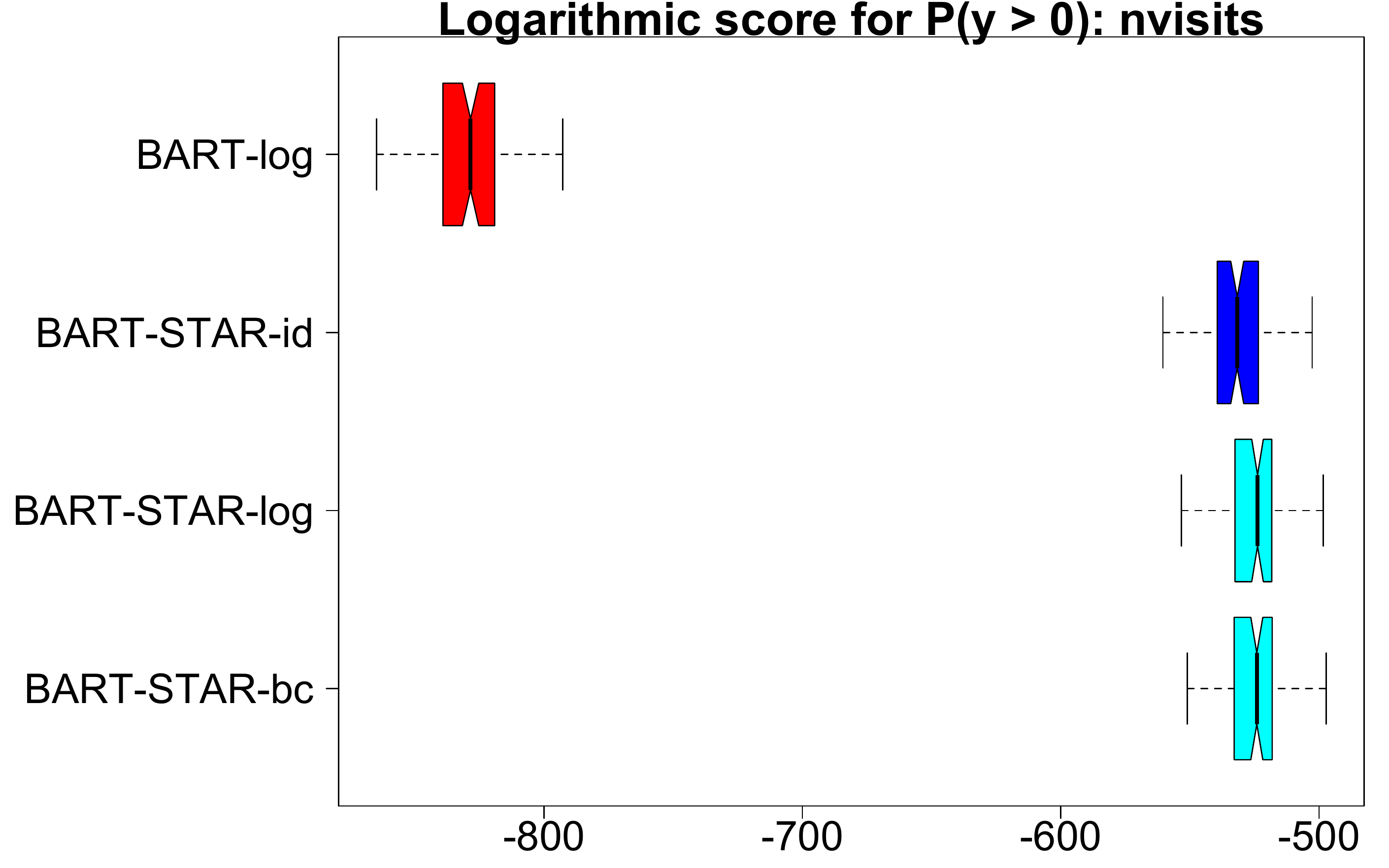}
\includegraphics[width=0.49\textwidth]{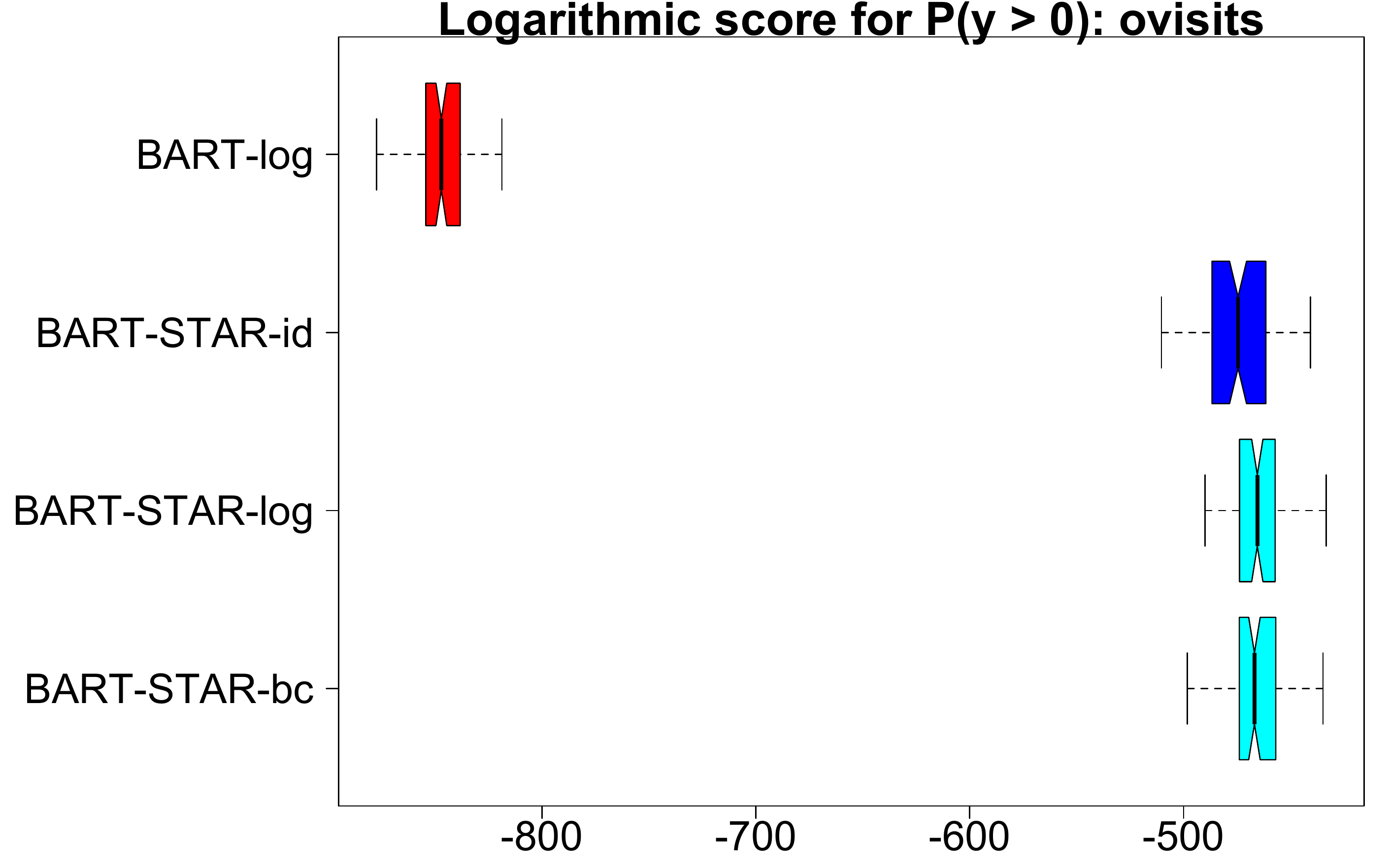}
\includegraphics[width=0.49\textwidth]{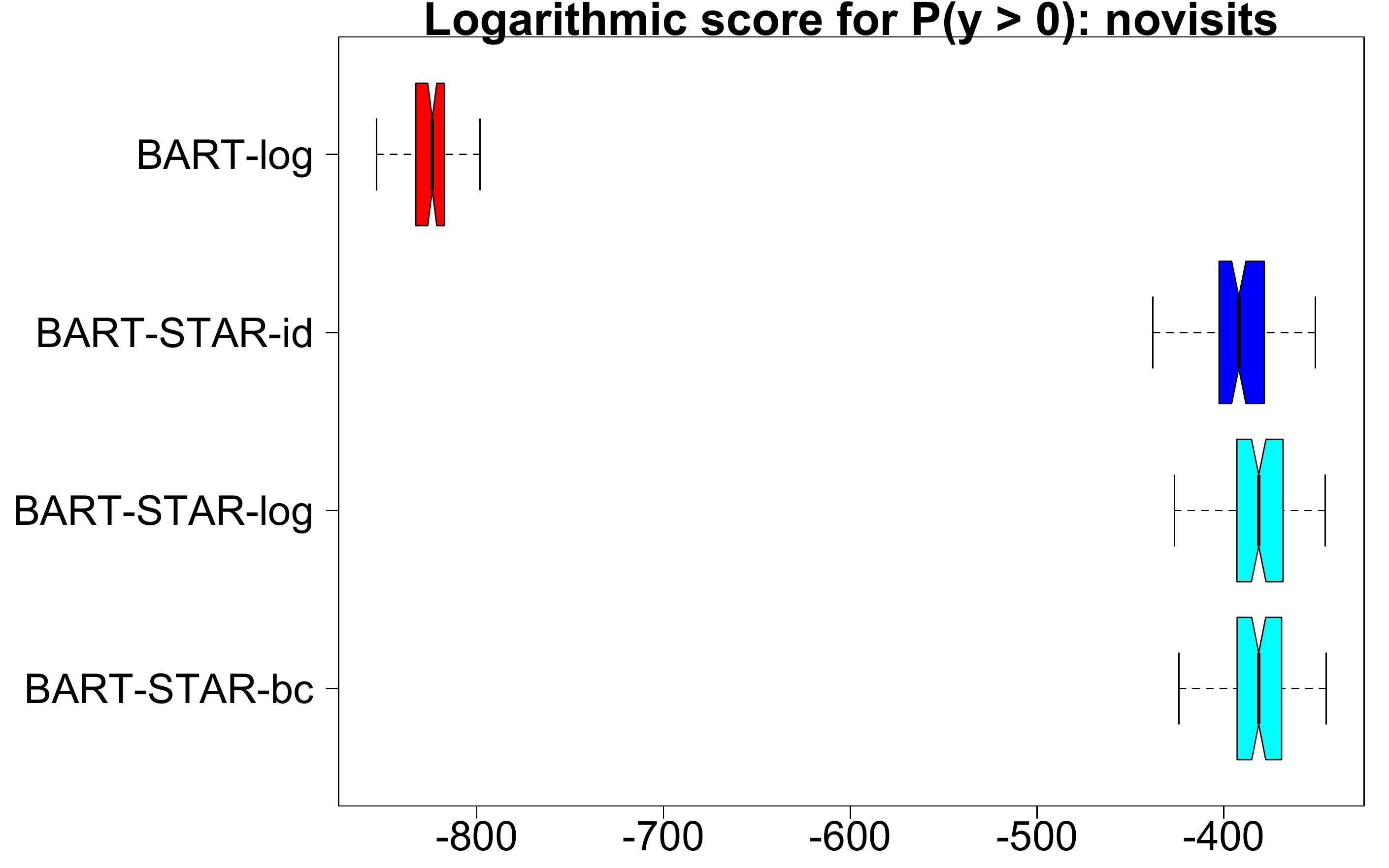}
\caption{\small Logarithmic scoring for estimation of $\mathbb{P}(\tilde y_i > 0 | y)$  across 100 test sets. Large scores indicate better performance. The \STAR models are clearly superior relative to \bartp-log, with transformed \STAR models (\STARp-log and \STARp-bc) offering the best performance overall. 
\label{fig:NMES-logarithmic}}
\end{center}
\end{figure}

Lastly, we consider a specific prediction task of interest: estimating the probability that an individual will utilize the healthcare system, $\mathbb{P}(\tilde y_i > 0 | y)$. We evaluate each method using logarithmic scoring on the event $\{\tilde y_i > 0\}$, which is a proper scoring rule for binary outcomes \citep{gneiting2007strictly}. The results are in Figure~\ref{fig:NMES-logarithmic}, where larger values indicate superior predictive accuracy. The \bartp-\STAR models which include both transformation and rounding (\bartp-\STARp-log and \bartp-\STARp-bc) decisively outperform the competitors, especially the continuous-valued model (\bartp-log). 

Based on multiple measures of predictive accuracy for each of four count-valued measures of healthcare utilization, the results from our out-of-sample comparison provide definitive confirmation of the predictive capabilities of \STAR models, and in particular \bartp-\STAR models which include \emph{both} transformation and rounding.

\section{Modeling the decline in Amazon river dolphins}\label{sec-dolphins}

The tucuxi dolphin (\emph{sotalia fluviatilis}) is a small river dolphin that inhabits the Amazon River. While the tucuxi dolphin population was once stable, the progression of habitat degradation, dolphin fishing, and other human interference has to led to increased concerns of population decline. To assess the validity of these concerns, \cite{da2018both} gathered data from 1994 to 2017 using multiple observers to search for tucuxi dolphins along a particular segment of the Amazon River. In addition to the number of tucuxi dolphins observed, the data include the water level (in meters), the number of observers present, and the date for each of $n=312$ surveys. While \cite{da2018both} fit a linear model to the logarithm of dolphin counts, we propose to leverage the \STAR modeling framework to investigate nonlinear effects and provide greater integer-valued distributional flexibility.

We use additive models to study  the yearly evolution of tucuxi dolphin counts,  which may be nonlinear, while adjusting for seasonal, water level, and observer effects. Specifically, for each survey we include the year (\texttt{year}), day-of-year (\texttt{doy}), and water level (\texttt{water}) as nonlinear predictors and the number of observers (\texttt{obs}) as a linear predictor. For comparisons, we implement a variety of Bayesian additive models: Gaussian additive models for the raw (\textsc{am}-id) and $\log(y+1)$ transformed data (\textsc{am}-log); \STAR models with identity (\textsc{am}-\STARp-id), unknown parametric (\textsc{am}-\STARp-bc), and unknown nonparametric (\textsc{am}-\STARp-np) transformations; and Poisson (\textsc{am}-Pois) and negative-binomial (\textsc{am}-NB) additive models (using \texttt{rstanarm}). For each method, we jointly evaluate the model performance and the computational efficiency: performance is measured using WAIC, while efficiency is reported as seconds per 10000 effective samples. In particular, we compute multivariate effective sample sizes \citep{vats2019multivariate} for 10 randomly sampled points from the posterior predictive distribution using the \texttt{mcmcse} package in \texttt{R} \citep{mcmcse}.  The results are in Table~\ref{table:tux-comp}. According to WAIC, the \STAR models with unknown transformation (\textsc{am}-\STARp-bc and \textsc{am}-\STARp-np) are strongly preferred, while \textsc{am}-NB is the closest competitor. However, the existing integer-valued additive models (\textsc{am}-Pois and \textsc{am}-NB) are noncompetitive in computational efficiency.

\begin{table}[ht]
\caption{WAIC and seconds per 10000 effective samples for additive models (dolphins data). \label{table:tux-comp} }
\resizebox{\textwidth}{!}{%
\begin{tabular}{rrrrrrrr}
  \hline
 & \textsc{am}-id & \textsc{am}-log & \textsc{am}-\STARp-id & \textsc{am}-\STARp-bc & \textsc{am}-\STARp-np & \textsc{am}-Pois & \textsc{am}-NB \\ 
  \hline
WAIC & 2171 & 1964 & 2000 & 1924 & {\bf 1916} & 2821 & 1931 \\ 
  Sec / 10000 ES & 32 & 30 & 52 & 67 & 95 & 843 & 295 \\ 
   \hline
\end{tabular}
}

{\raggedright 
     \vspace{1ex}
Multivariate effective sample sizes are computed for 10 randomly selected points from the posterior predictive distribution. Both transformation \emph{and} rounding via \STAR are essential to achieve competitive performance. The \STAR MCMC performance is strong, especially relative to the Poisson and negative-binomial models (using \texttt{rstanarm}). \par}
\end{table}


Proceeding with \textsc{am}-\STARp-np, which is selected by WAIC, we plot posterior expectations and credible intervals for each $f_j$ in Figure~\ref{fig:am-tux}. MCMC diagnostics for $f_j$ show exceptional  mixing with no lack of convergence (see Appendix~\ref{app:dolphins}). 
The \texttt{doy} plot suggests a seasonal pattern, while the \texttt{water} plot exhibits an approximately quadratic effect. Most importantly, the \texttt{year} plot shows a near linear decline in tuxucis dolphins from 1994-2017, which interestingly has leveled off since 2013. These findings are partially consistent with the results of \citet{da2018both} which, assuming only a linear model, also report a significant decrement of dolphins since 1994. Posterior predictive diagnostics are in Appendix~\ref{app:dolphins}, and indicate clear improvements in fit for \textsc{am}-\STARp-np relative to models that exclude transformation (\textsc{am}-\STARp-id) or rounding (\textsc{am}-log).

\begin{figure}[h]
\begin{center}
\includegraphics[width=.32\textwidth]{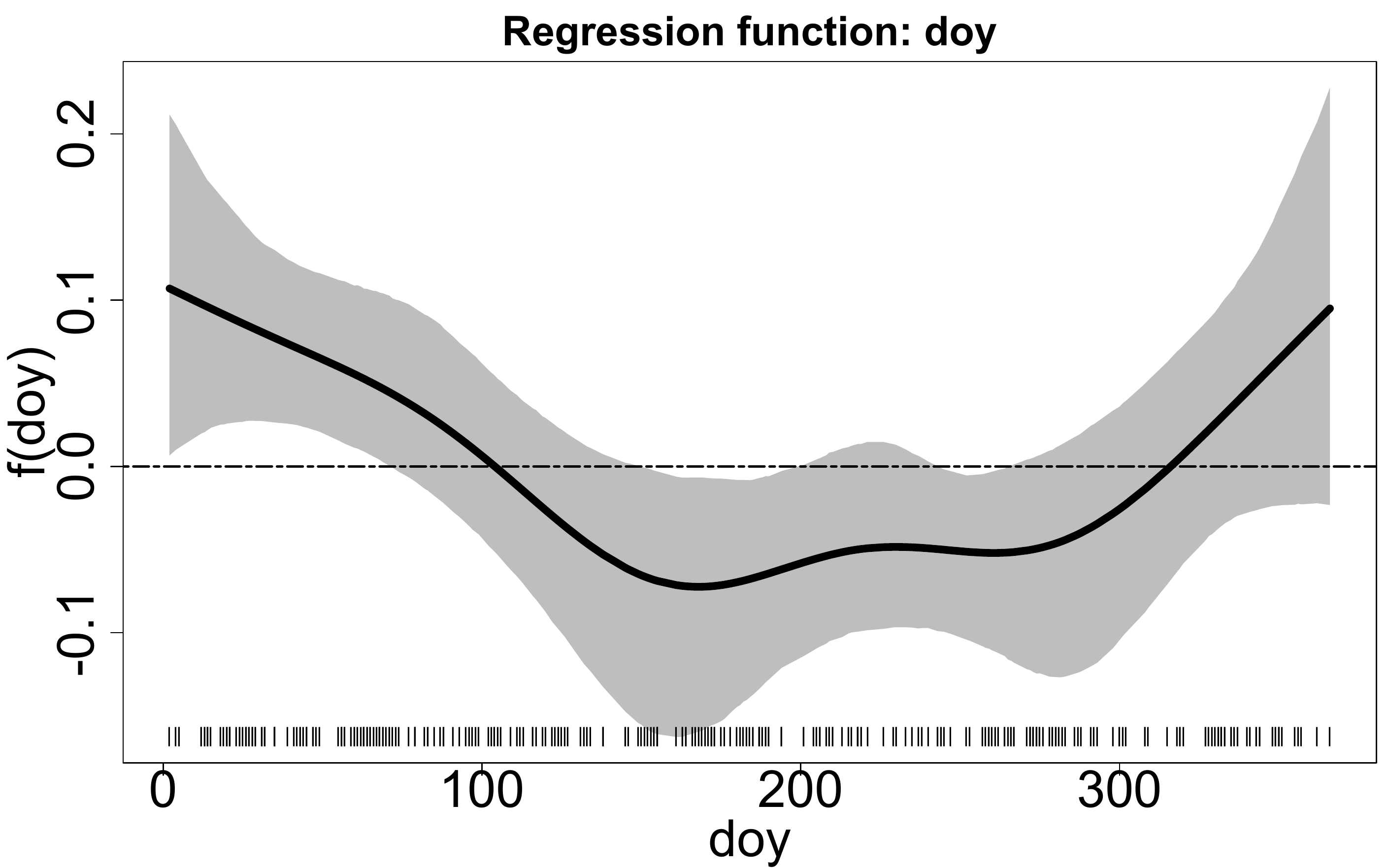}
\includegraphics[width=.32\textwidth]{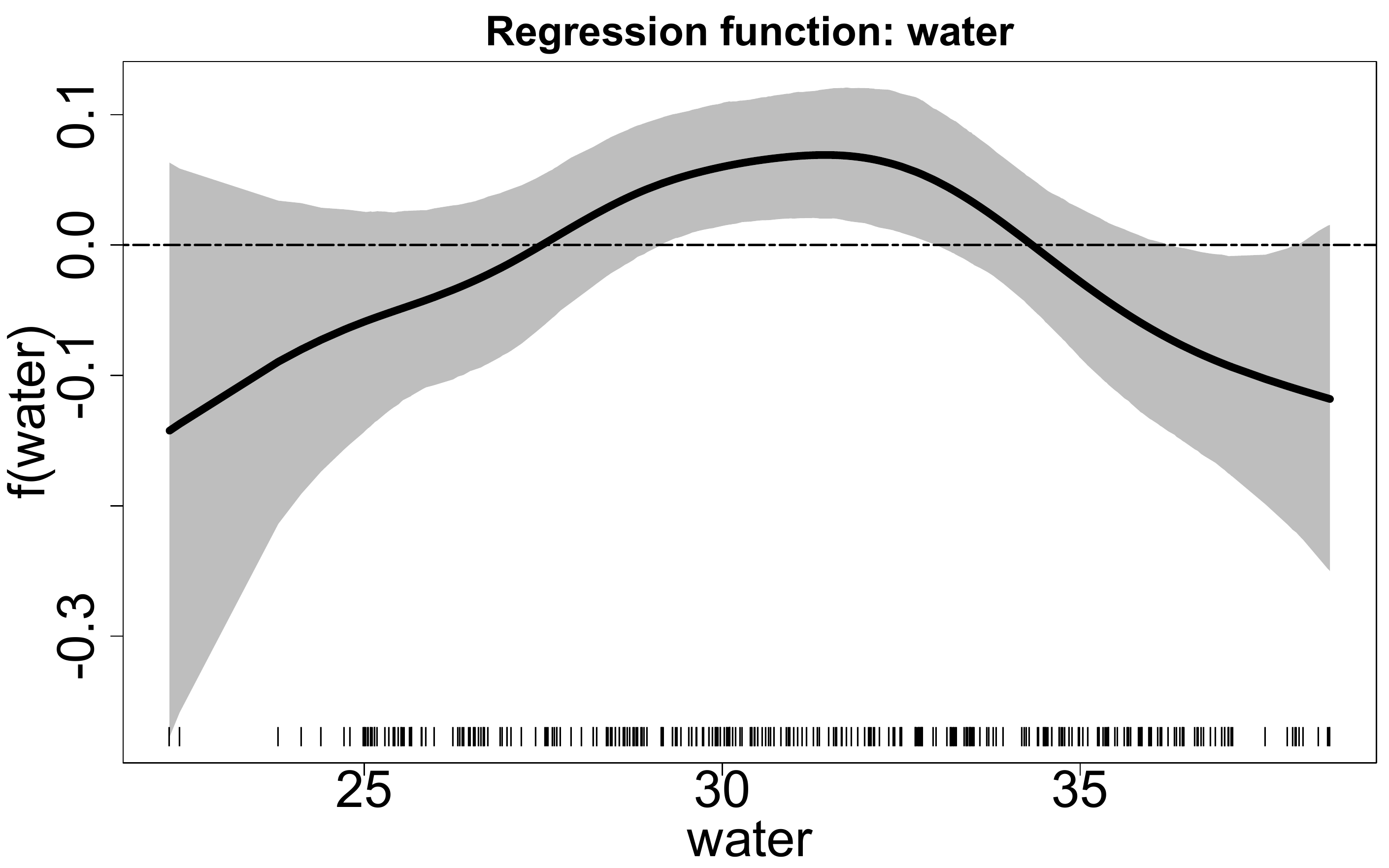}
\includegraphics[width=.32\textwidth]{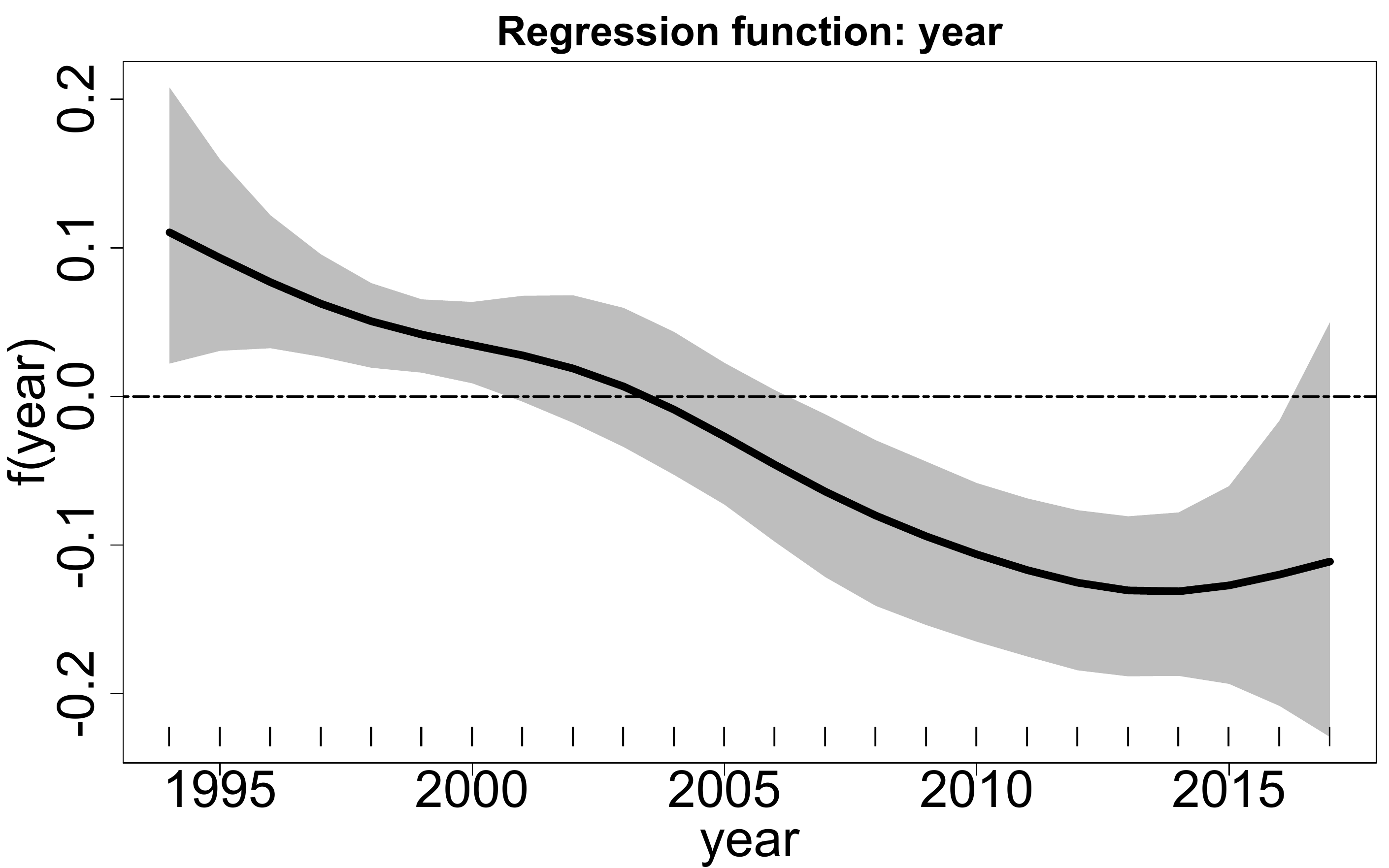}
\caption{\small Posterior expectation and 95\% pointwise credible intervals for each $f_j(v_j)$ under \textsc{am}-\STARp-np for the dolphins data. The tick marks indicate the observation points for each predictor. 
\label{fig:am-tux}}
\end{center}
\end{figure}

\section{Discussion}\label{discuss}
\STAR processes provide a mathematically elegant and empirically
successful framework to model count and integer-valued data. The
approach seamlessly adapts state-of-the-art continuous data models and
algorithms to the integer-valued data setting, and thus it offers
remarkable modularity both in terms of model specification and
computation, while providing ease of implementation and interpretability for 
practitioners. By incorporating known, unknown parametric, and unknown
nonparametric transformations, \STAR processes provide varying degrees
of distributional flexibility, and are able to account for important
distributional features such as zero-inflation, bounded or censored
data, and over- or underdispersion. Empirically, \STAR
processes demonstrate goodness-of-fit, out-of-sample point and interval predictive accuracy, reliable inference, and computational scalability. 

In addition to the healthcare utilization and animal abundance datasets considered here, we provide further empirical comparisons on three additional datasets in Appendix~\ref{app:emp}. Among \STARp, Gaussian, Poisson, and negative-binomial models, the  \STAR additive and \bartp-\STAR models with unknown transformations consistently provide the best performance according to WAIC.

A variety of promising extensions exist for \STARp. The modeling and computational modularity of \STAR suggest that new multivariate, functional, and time series models may be developed for integer-valued data. Furthermore, \STAR is capable of modeling rounded data, which is ubiquitous in practice yet rarely considered in modern statistical and machine learning methods. Lastly, the \STAR model \eqref{round}-\eqref{transform} does not strictly require a Bayesian modeling approach, and may be adapted for classical estimation and inference.

\appendix

\section{Supplemental results} 

\subsection{Evaluating point accuracy for synthetic data}\label{app:sims}
To accompany the WAIC comparisons from the main paper, we evaluate each method for point estimation accuracy. Specifically, we are interested in estimating the conditional expectation of the observed data, $\lambda_i^*(x)$. For an estimator $\hat y_i(x)$, we compute the root mean squared error $\mbox{RMSE} = \sqrt{\sum_{i=1}^n \big\{\lambda_i^*(x) - \hat y_i(x)\big\}^2}$. The fitted values for \STAR are computed using the conditional expectation of $y$ at $x \in \mathcal{X}$, that is
\begin{equation}\label{cond-exp}
 \mathbb{E}\{y(x)\} = \sum_{j=0}^\infty j \mathbb{P}\{y(x) = j\} \approx \sum_{j=1}^{J(x)} j \mathbb{P}\{y(x) = j\},
\end{equation}
where $\mathbb{P}\{y(x) = j\}$ is the \STAR probability mass function and $J(x)$ is a finite truncation. Since $\mathbb{P}\{y(x) = j\}$ depend on the parameters $\theta$ in $\Pi_\theta$, the posterior distribution of \eqref{cond-exp}
may be computed by evaluating \eqref{cond-exp} for each draw of $\theta$ in the MCMC algorithm. Conservatively, we select $J(x)$ to be the 99.99th quantile of the distribution of $y(x)$ pointwise for each $x$, which is easily computable as $h\left[g^{-1}\left\{z_q^*(x)\right\}\right]$ where $z_q^*(x)$ is the $q$th quantile of $\Pi_\theta$. The point estimate is computed as the posterior expectation of \eqref{cond-exp}. 

\begin{figure}[h]
\begin{center}
\includegraphics[width=0.49\textwidth]{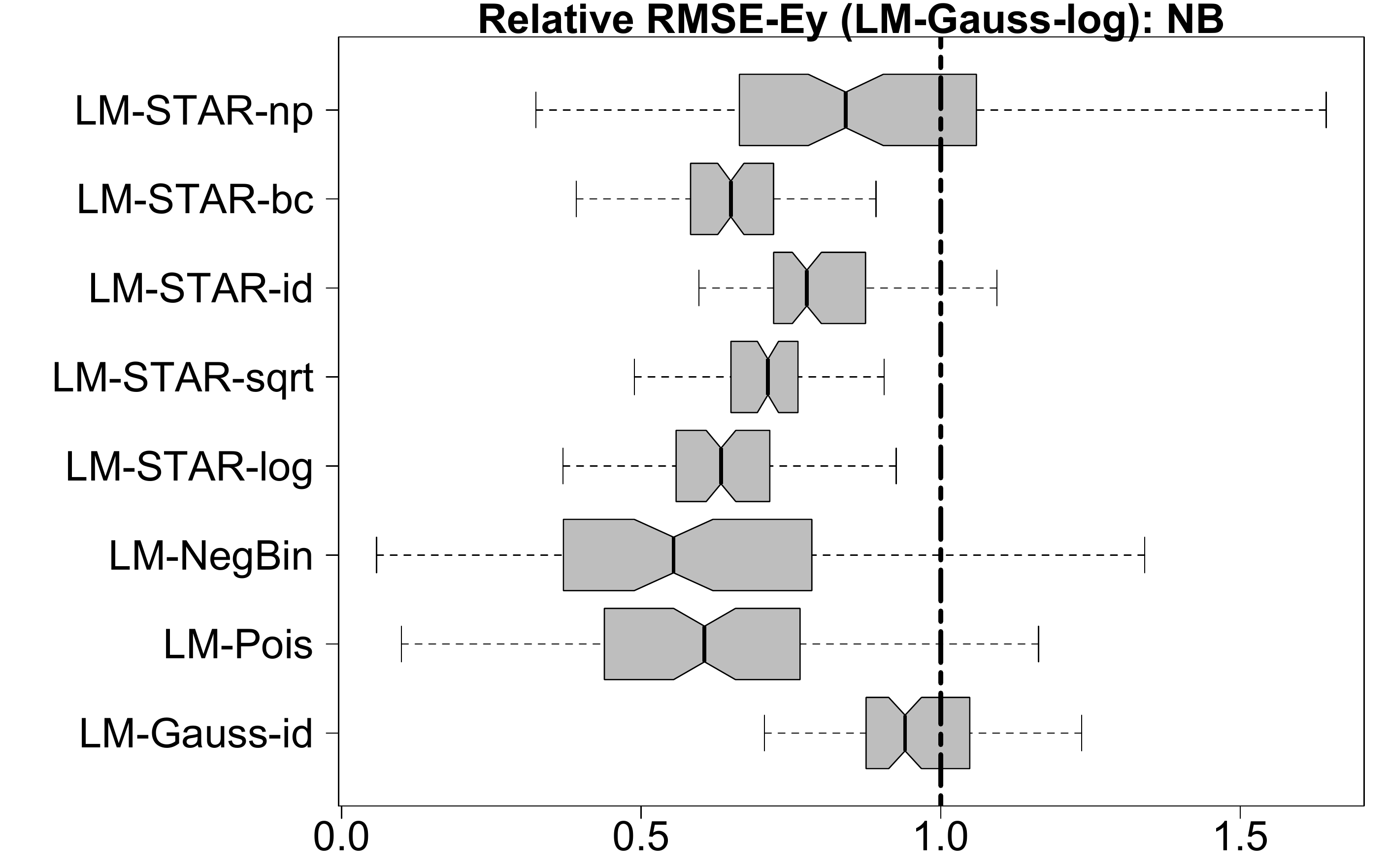}
\includegraphics[width=0.49\textwidth]{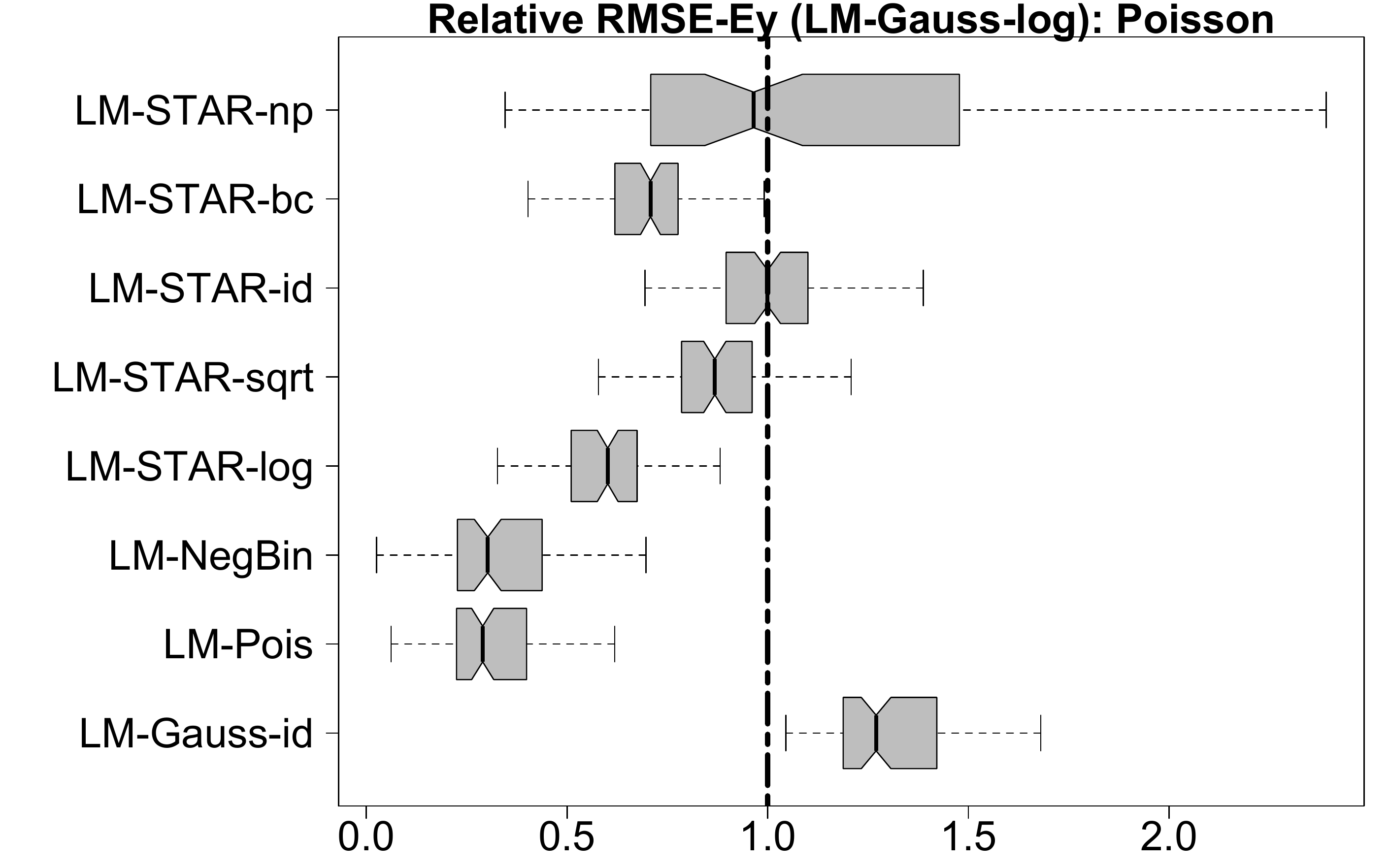}
\caption{\small Relative RMSE under various distributions. Preferred models have smaller values, and models with values less than 1.0 are preferred to \textsc{lm}-log. As expected, the \textsc{lm}-NegBin performs well, since it closely matches the data-generating process. Notably, the \STAR models are highly competitive, and clearly superior to the Gaussian models, especially \textsc{lm}-\STARp-bc and \textsc{lm}-\STARp-log. 
\label{fig:sim-lm-rmse}}
\end{center}
\end{figure}

\begin{figure}[h]
\begin{center}
\includegraphics[width=0.49\textwidth]{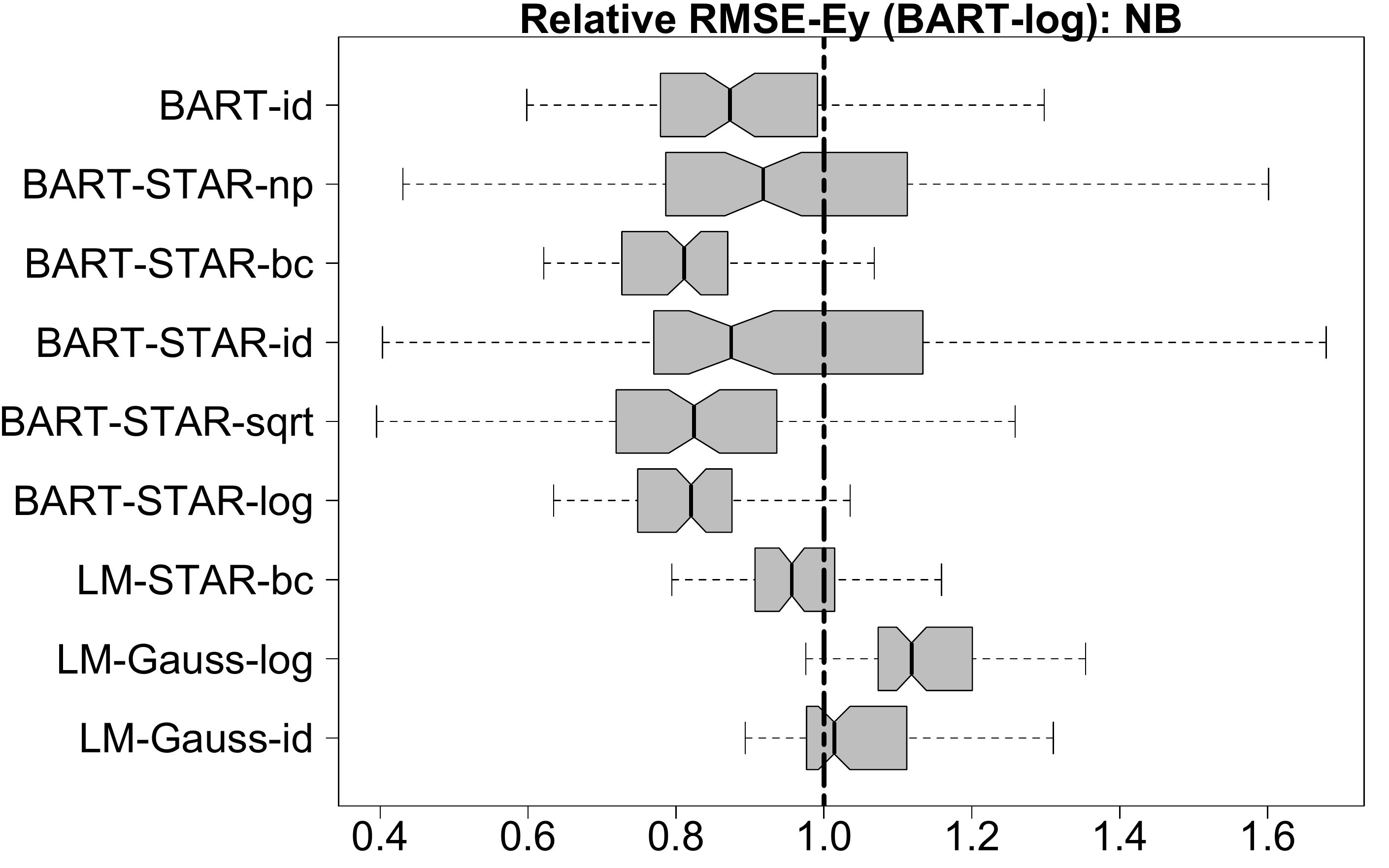}
\includegraphics[width=0.49\textwidth]{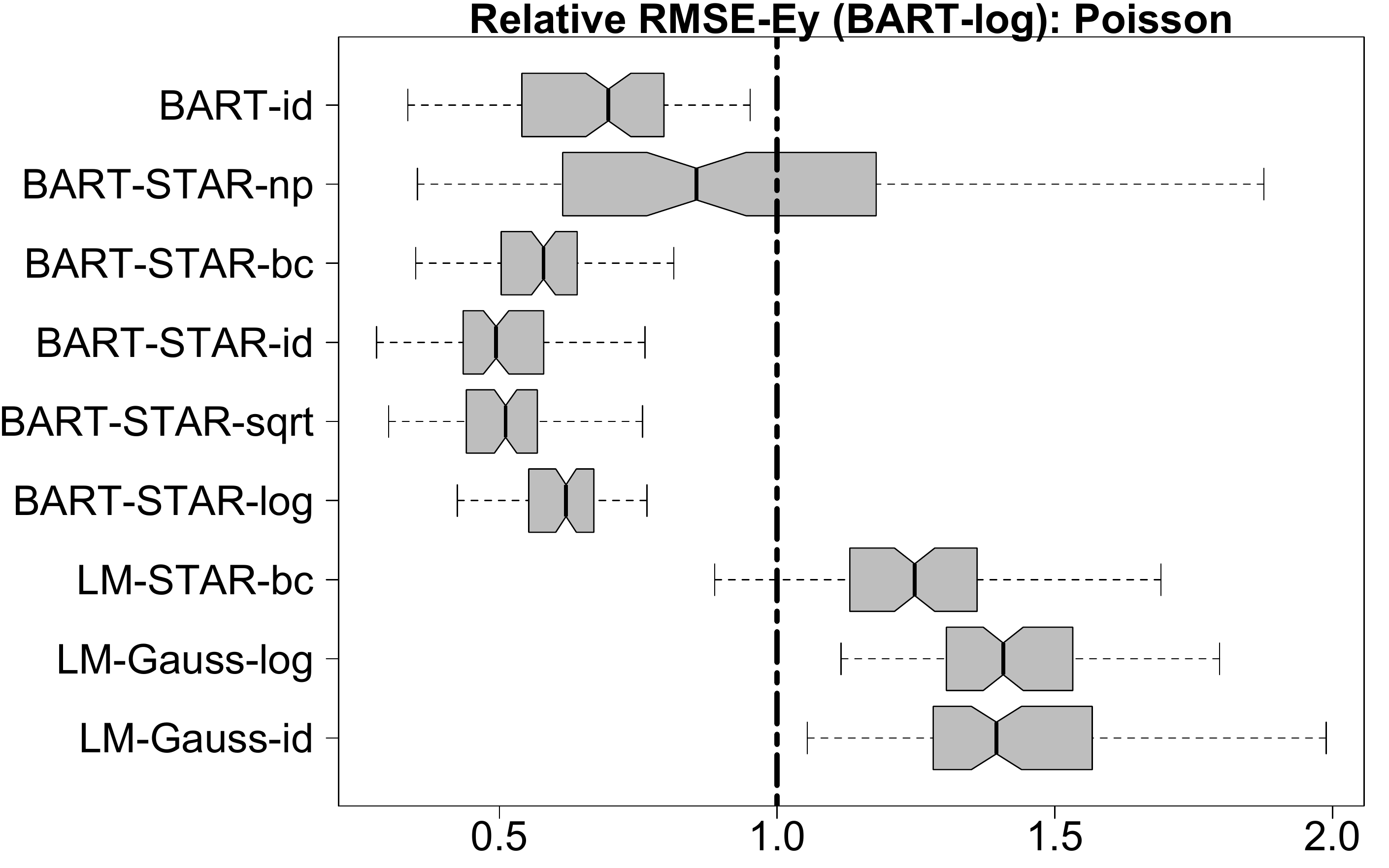}
\caption{\small Relative RMSE under various distributions. Preferred models have smaller values, and models with values less than 1.0 are preferred to \bartp-log. 
\label{fig:sim-bart-rmse}}
\end{center}
\end{figure}

Figures~\ref{fig:sim-lm-rmse}~and~\ref{fig:sim-bart-rmse} depict the relative RMSEs across simulated data sets for the linear and nonlinear simulation designs in Sections~4.1~and~4.2, respectively, defined as the ratio between the RMSE of the generic model over the RMSE for a baseline method, and specifically the \textsc{lm}-log method which represents the common approach of modeling (log-) transformed counts using Gaussian models. Relative RMSE standardizes model performance across simulated datasets: methods with a relative RMSE less than 1.0 demonstrate  superior point estimation relative to the baseline method. As in Section~4, we find that \STARp-log and \STARp-bc are consistently competitive and outperform other methods. Interestingly, \STARp-np is much less competitive in RMSE than in WAIC, which suggests that the additional distributional flexibility acquired by modeling $g$ nonparametrically does not necessarily imply more accurate point estimation.

\subsection{Model and MCMC diagnostics for the dolphins data}\label{app:dolphins}

\begin{figure}[h]
\begin{center}
\includegraphics[width=1\textwidth]{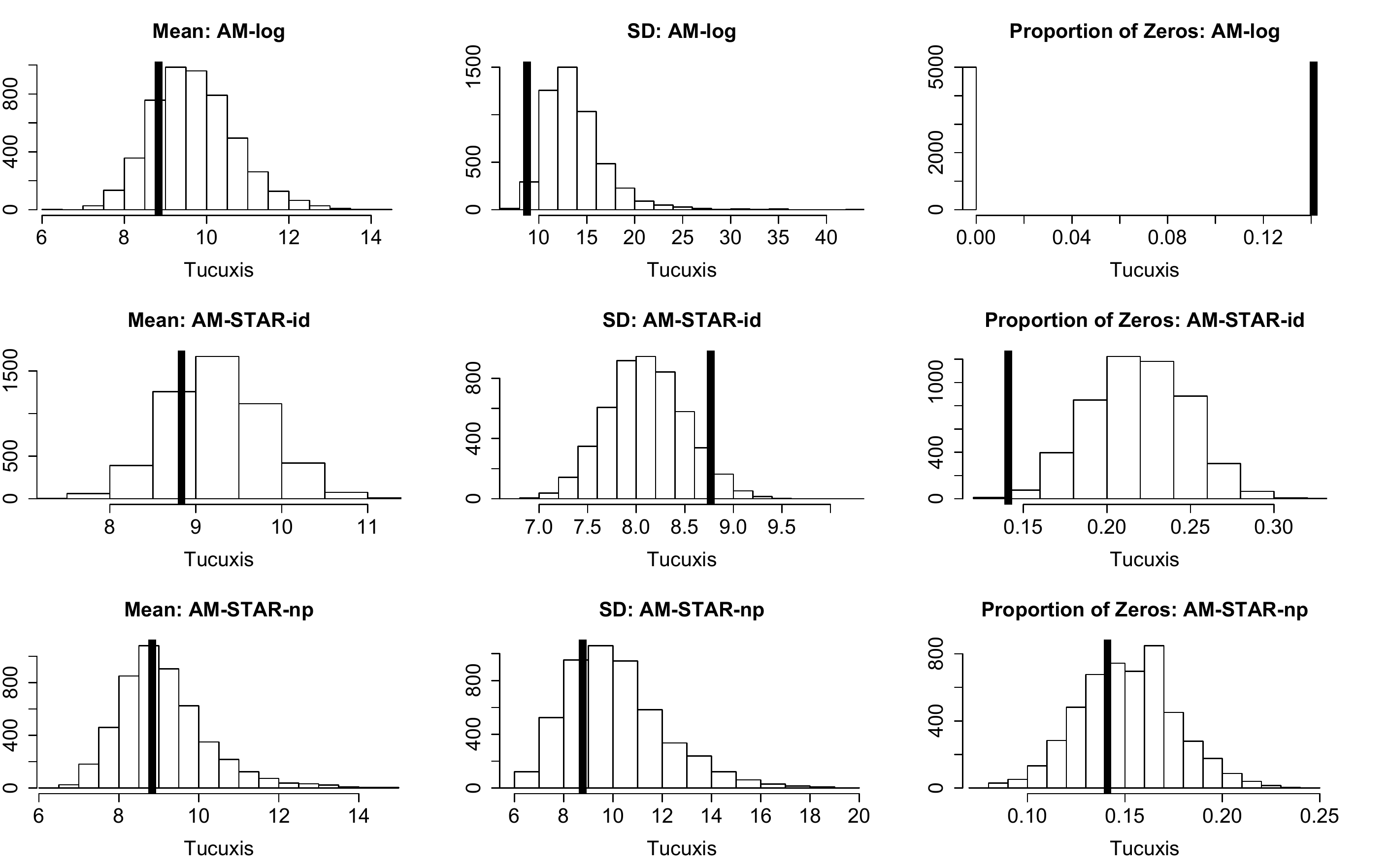}
\caption{\small Posterior predictive diagnostics for \textsc{am}-log (top row),  \textsc{am}-\STARp-id (middle row), and \textsc{am}-\STARp-np (bottom row). The  mean (left), standard deviation (center), and proportion of zeros (right) were computed for each posterior predictive simulated dataset (histograms) and the observed data $y$ (vertical lines). Only the model including both transformation \emph{and} rounding (\textsc{am}-\STARp-np) is adequate by these measures for the dolphins data. 
\label{fig:postpred-am-tux}}
\end{center}
\end{figure}

Posterior predictive diagnostics for additive models fit to the tucuxis dolphins data are in Figure~\ref{fig:postpred-am-tux}. The additive \STARp-np model is adequate for the data, while the models which lack either rounding or transformation are incapable of capturing distributional features, including the variability and the proportion of zeros. 

\begin{figure}[h]
\begin{center}
\includegraphics[width=1\textwidth]{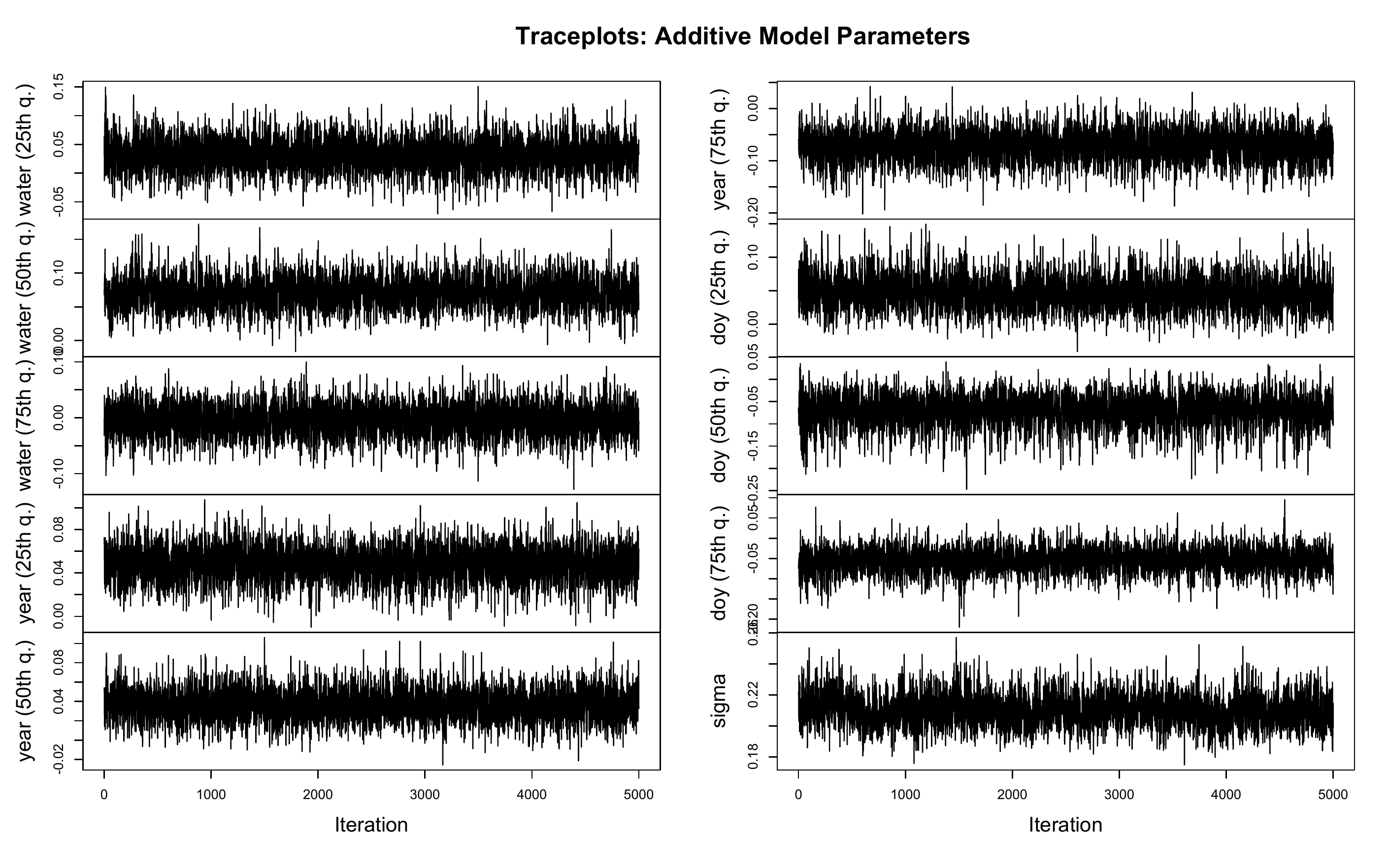}
\caption{\small Traceplots for $f_j(v_j)$ and $\sigma$ under the additive \STARp-np model for the dolphin data, where the functions $f_j$ are evaluated at the 25th, 50th, and 75th sample quantiles of each $\{v_{i,j}\}_{i=1}^n$. The MCMC chain consisted of 5000 iterations (after discarding a burn-in of 5000 and retaining every 3rd sample). Effective sample sizes for all $f_j(v_j)$ exceeded 2000. 
\label{fig:trace-dolphins}}
\end{center}
\end{figure}

The MCMC convergence of the additive \STARp-np model is assessed via traceplots in Figure~\ref{fig:trace-dolphins}. The traceplots indicate no lack of convergence and demonstrate exceptional mixing: effective sample sizes for all $f_j(v_j)$ exceed 2000.

\subsection{Supplemental empirical examples}\label{app:emp}
For further models comparisons, we apply the linear, additive, and \bart models to several additional datasets, and again consider \STARp, Gaussian, Poisson, and negative-binomial distributions. For the additive models, each continuous variable (i.e., variables with at least 10 unique observation points) is modeled nonlinearly.

\subsubsection*{Ships Data}
The \texttt{ships} data, available in the \texttt{MASS} package in \texttt{R}, provides the number of damage incidents due to waves for $n = 34$ cargo-carrying vessels, as well as ship type (A-E), year of construction (1960-1964, 1965-1969, 1970-1974, or 1975-1979), the period of construction (1960-1974 or  1975-1979), and the aggregated months of service (ranging from 0 to 44882). We model the ship type, year of construction, and period of construction as factors, and center and scale the service variable. The data were analyzed in \cite{mccullagh1989generalized} using a quasi-Poisson regression model to account for observed overdispersion, and subsequently re-analyzed in \cite{mallick1994generalized} using a Poisson regression model with unknown link function, which suggests that additional distributional flexibility in the regression model may be important.



\subsubsection*{Roaches Data}
\cite{gelman2006data} consider a study of pest management for eliminating cockroaches in city apartments. The response variable, $y_i$, is the number of roaches caught in traps in apartment $i$, with $i=1,\ldots, n = 262$. A pest management treatment was applied to a subset of 158 apartments, with the remaining 104 apartments receiving a control. Additional data are available on the pre-treatment number of roaches, whether the apartment building is restricted to elderly residents, and the number of days for which the traps were exposed. A notable feature of the data is zero-inflation: $y_i = 0$ for 94 (36\%) of the apartments.

\subsubsection*{Highway Data}
The \texttt{Highway} data, available in the \texttt{carData} package in \texttt{R},  consists of the 1973 accident rate per million vehicle miles on $n =39$ large sections of Minnesota highway. Important predictors include the number of access points per mile, the speed limit, the width of the outer shoulder on the roadway (in feet), and the number of signals per mile of roadway, among others. We consider the accident rate per 10,000 miles, which is the smallest rate for which the observations $y_i$  are integer-valued. A notable feature of these data are that, despite being (scaled) accident counts, no two observations  $y_i$ and $y_j$ are equal, and the counts themselves are large, ranging from 161 to 923. Therefore, it is unclear \emph{a priori} whether an integer-valued model is necessary or advantageous.

\subsubsection*{Results}
The WAICs for the supplementary datasets are reported in Table~\ref{table:allApps}. The \STAR models consistently perform well across all datasets, and in particular \STARp-bc and \STARp-np. Notably, \STAR provides the best linear, additive, and \bart models for all datasets with the exception of the \texttt{Highway} data, for which all \bart models perform similarly. Interestingly, additive \STAR models are preferred for both the \texttt{ships} data and the \texttt{Highway} data, while \bartp-\STAR is slightly preferred to the additive \STAR model for the \texttt{roaches} data.

\begin{table}[ht]
\caption{WAIC for the \texttt{ships}, \texttt{roaches}, and \texttt{Highway} datasets.  \label{table:allApps}}

\small \centering
\resizebox{\textwidth}{!}{%
\begin{tabular}{rrrrrrrrrrr}

  \hline
& Model & Gauss-id & Gauss-log & \STARp-log & \STARp-sqrt & \STARp-id & \STARp-bc & \STARp-np & Pois & NB \\ 
  \hline

\multirow{3}{*}{\texttt{ships}}
& \textsc{lm} & 292 & 204 & 204 & 195 & 220 & 196 & {\bf 194} & 239 & 203 \\ 
  & \textsc{am} & 240 & 188 & 193 & 173 & 182 & 175 & {\bf \underline{171}} & - & - \\ 
 &  \bart & 242 & 204 & 200 & {\bf 182} & 191 & 196 & 187 & - & - \\  
   \hline
   
\multirow{3}{*}{\texttt{roaches}}
& \textsc{lm} & 2722 & 1886 & 1791 & 1772 & 1952 & {\bf 1756} & 1759 & 12565 & 1793 \\ 
  & \textsc{am} & 2711 & 1838 & 1740 & 1732 & 1928 & {\bf 1710} & 1729 & - & - \\ 
 &  \bart & 2700 & 1833 & 1736 & 1732 & 1927 & {\bf \underline{1708}} & 1719 & - &  - \\ 
   \hline

\multirow{3}{*}{\texttt{Highway}}
& \textsc{lm} & 495 & 481 &  481 & 484 & 496 & 486 & {\bf 479} & 1587 & 496 \\
  & \textsc{am} & 487 & 465 & 465 & 473 & 484 & 478 & {\bf \underline{460}} & - & - \\
 &  \bart & 466 & {\bf 463} & 465 & 465 & 469 & 476 & 464 & - &  -\\ 
   \hline

     \hline
   
\end{tabular}
}
\raggedright{
     \vspace{1ex}

The best method (lowest WAIC) for each model class is in bold, and the best overall model for each outcome is underlined. The \STAR models dominate, with the additive and \bart models typically outperforming the linear models. 
}
\end{table}

\bibliographystyle{apalike}
\bibliography{refs}

\end{document}